\documentclass[a4paper,11pt]{article}
\pdfoutput=1 

\usepackage{jheppub} 

\usepackage[T1]{fontenc} 
\usepackage{calc,dsfont,array,amsmath,amsfonts,graphicx,amsthm,amssymb,slashed}

\newcommand{\g}[1]{\gamma_{#1}} 
\renewcommand{\l}{\left}
\renewcommand{\r}{\right}

\newcommand{\Tr}{\mathrm{Tr}}
\newcommand{\diag}{\mathrm{diag}}  
\newcommand{\bra}[1]{\left< #1 \right|} 
\newcommand{\ket}[1]{\left| #1 \right>} 

\newcommand{\order}[1]{\mathcal{O}\l({#1}\r)}

\newcommand{\SU}[1]{\mathrm{SU}\l(#1\r)}

\newcommand{\expect}[1]{\left\langle #1 \right\rangle}
\newcommand{\A}{\mathcal{A}}
\newcommand{\B}{\mathcal{B}}
\newcommand{\C}{\mathcal{C}}
\newcommand{\chitop}{\chi_{\rm top}}
\newcommand{\obs}{\mathcal{O}}

\newcommand{\RM}{\mathbb{R}_M}
\newcommand{\psibar}{\overline{\psi}}
\newcommand{\chibar}{\overline{\chi}}
\newcommand{\stat}{\mathrm{stat}}
\newcommand{\sys}{\mathrm{sys}}

\title{\boldmath Non-perturbative Test of the Witten-Veneziano Formula from Lattice QCD}

\author[a,b,c]{Krzysztof Cichy,}
\author[b,d]{Elena Garcia-Ramos,}
\author[b]{Karl Jansen,}
\author[e]{Konstantin Ottnad,}
\author[e]{Carsten Urbach}

\affiliation[a]{Goethe-Universit\"at, Institut f\"ur Theoretische Physik, Max-von-Laue-Stra\ss e 1, D-60438 Frankfurt a.M., Germany}
\affiliation[b]{NIC, DESY, Platanenallee 6, D-15738 Zeuthen, Germany}
\affiliation[c]{Adam Mickiewicz University, Faculty of Physics, Umultowska 85, 61-614 Poznan, Poland} 
\affiliation[d]{Humboldt Universit\"at zu Berlin, Newtonstr. 15, D-12489 Berlin, Germany}
\affiliation[e]{Institut f\"ur Strahlen- und Kernphysik (Theorie), Nussallee 14-16 and Bethe Center for Theoretical Physics, Nussallee 12, Universit\"at Bonn, D-53115 Bonn, Germany}

\emailAdd{krzysztof.cichy@desy.de}
\emailAdd{elenagr@ifh.de}
\emailAdd{Karl.Jansen@desy.de}
\emailAdd{ottnad@hiskp.uni-bonn.de}
\emailAdd{urbach@hiskp.uni-bonn.de}

\collaborationImg{\includegraphics[width=0.125\textwidth]{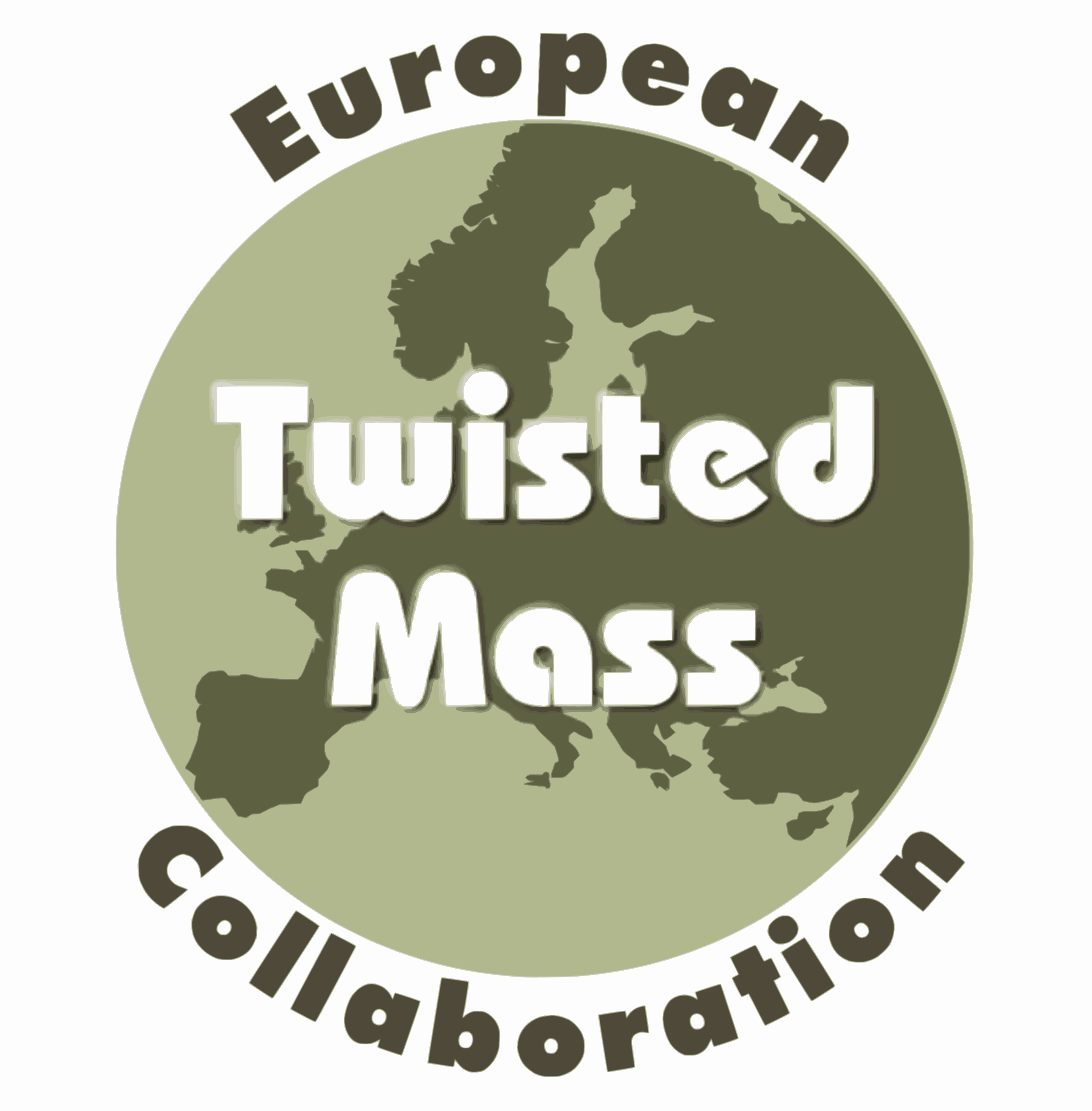}}
\collaboration{The ETM collaboration}
\preprint{DESY 15-051, SFB/CPP-14-123}
\arxivnumber{1504.07954}
\graphicspath{{plots/}}

\abstract{We compute both sides of the Witten-Veneziano formula using lattice
techniques. For the one side we perform dedicated quenched simulations
and use the spectral projector method to determine the topological
susceptibility in the pure Yang-Mills theory. The other side we determine
in lattice QCD with $N_f=2+1+1$ dynamical Wilson twisted mass fermions
including for the first time also the flavour singlet decay constant.
The Witten-Veneziano formula represents a leading order expression in the framework of chiral perturbation theory
and we also employ leading order chiral perturbation theory 
to relate the flavor singlet decay constant to the relevant decay constant parameters in the quark 
flavor basis and flavor non-singlet decay constants. After taking the continuum and the SU$(2)$ chiral limits we compare
both sides and find good agreement within uncertainties.}

\begin{document} 
\maketitle
\flushbottom
\section{Introduction}

The Witten-Veneziano formula \cite{Witten:1979vv,Veneziano:1979ec} was first derived by Witten for massless quarks
\begin{equation}
 \label{eq:W-Vformula_chiral}
 \mathring{M}_{\eta^{\prime}}^2=\frac{4N_f}{f_0^2}\chi_{\infty}\,.
\end{equation}
It aims at providing an explanation for the unexpectedly large mass of the $\eta'$ meson, by relating its mass in the chiral limit $\mathring{M}_{\eta^{\prime}}$ to non-trivial topological fluctuations of the gauge fields, which are encoded in the topological susceptibility computed in pure Yang-Mills (YM) theory $\chi_\infty$. Furthermore, the formula involves the singlet decay constant $f_0$ and the number of quark flavors $N_f$.

In order to obtain Eq.~(\ref{eq:W-Vformula_chiral}) one considers the limit of a large number of colors $N_c$, for which simplifications of the theory occur that allow one to address a variety of problems which are otherwise impossible to investigate. In particular, in the 't Hooft limit ($N_c\rightarrow \infty$, while $g^2N_c$ and $N_f$ are kept fixed, with $g$ denoting the gauge coupling) the anomalously broken axial $U(1)$ symmetry is restored and the $\eta'$ becomes a Goldstone boson, i.e. $\mathring{M}_{\eta^{\prime}}\rightarrow 0$ \cite{'tHooft:1973jz,Witten:1979vv}. The formula itself is valid up to corrections of $\obs(1/N_c^2)$ while $\mathring{M}_{\eta^{\prime}}^2$ and $1/f_0^2$ vanish as $\obs(1/N_c)$. However, the topological susceptibility which appears on the right-hand side of Eqs.~(\ref{eq:W-Vformula_chiral}) remains finite in the large $N_c$ limit, i.e. $\chi_{\infty}=\mathcal{O}(1)$, such that in the $N_c\rightarrow\infty$ limit both sides of Eq.~(\ref{eq:W-Vformula_chiral}) vanish. The question remains whether $N_c=3$ in QCD is large enough in practice to sufficiently suppress corrections to this formula to higher order in $1/N_c$.

An alternative way to derive the Witten-Veneziano formula is to expand the anomalous flavor-singlet Ward-Takahashi identities of the theory order by order in $u=N_f/N_c$ around $u=0$ \cite{Veneziano:1979ec,Veneziano:1980xs}. Again, the Witten-Veneziano formula corresponds to the lowest order relation in this expansion. Besides, we remark that for lattice QCD it is also possible to obtain an unambiguous, theoretical sound implementation of the Witten-Veneziano formula through the study of anomalous flavor-singlet Ward-Takahashi identities in the limit $u\rightarrow 0$ \cite{Giusti:2001xh,Giusti:2001km}.

However, the very first attempts to compute that topological susceptibility in YM theory date back to the early 80s \cite{DiVecchia:1981qi,DiVecchia:1981hh} and a first reasonable number was found in \cite{DiGiacomo:1989id}.

Moving away from the chiral limit leads to corrections to the formula which are linear in the quadratic meson masses
\begin{equation}
 \label{eq:W-Vformula}
 \frac{f_0^2}{4N_f}(M_\eta^2+M_{\eta'}^2-2M_K^2)=\chi_{\infty} \,.
\end{equation}
This result was first derived in \cite{Veneziano:1979ec} from the aforementioned expansion in $u$. In the above expression terms have been reshuffled compared to Eq.~(\ref{eq:W-Vformula_chiral}) to isolate the topological susceptibility in the YM theory on the r.h.s., whereas the l.h.s. of the formula contains only quantities that need to be computed in full QCD. 

From a modern point of view, the Witten-Veneziano formula can also be derived from effective field theory employing a combined power counting scheme in quark masses $m_q$, momenta $p$ and $1/N_c$, given by $m_q=\mathcal{O}(\delta)$, $p^2 \mathcal{O}(\delta)$ and $1/N_c = \mathcal{O}(\delta)$, where $\delta$ is a small expansion parameter \cite{Gasser:1984gg, Kaiser:1998ds, Feldmann:1999uf, Kaiser:2000gs}. In principle, this approach allows one to systematically calculate higher order corrections to the standard form of Eq.~(\ref{eq:W-Vformula})
\begin{equation}
 M_\eta^2+M_{\eta'}^2-2M_K^2=\frac{4N_f}{f_0^2}\chi_{\infty} \,,
\end{equation}
which represents the leading order expression with respect to the expansion in $\delta$ used in chiral perturbation theory ($\chi$PT). Note that in general $f_0\neq f_\pi$ holds, as the singlet decay constant $f_0$ in Eqs.~(\ref{eq:W-Vformula_chiral}), (\ref{eq:W-Vformula}) becomes equal to the octet decay constant $f_8$ only when simultaneously taking the chiral limit and dropping all corrections in $1/N_c$ \cite{Kaiser:1998ds}. Since on the lattice we do not work in the octet-singlet basis, we cannot compute $f_0$ directly but need to apply leading order chiral perturbation theory in order to relate $f_0$ to the corresponding decay constant parameter in the quark flavor basis and the relevant flavor non-singlet decay constants $f_\pi$ and $f_K$.

In this paper, we compute both the topological susceptibility in the pure YM theory (or quenched QCD) and the meson masses and the singlet decay constant $f_0$ in full QCD. After discussing the lattice actions, we will first discuss the determination of $\chi_\infty$ using the so-called spectral projector method~\cite{LuscherGiusti} based on dedicated quenched simulations\footnote{For early attempts to compute the topological susceptibility in YM theory we refer to \cite{DiVecchia:1981qi,DiVecchia:1981hh}, while a first reasonable number was found in \cite{DiGiacomo:1989id}.}. Thereafter, we discuss the determination of $M_\eta, M_{\eta'}, M_K$ and in particular $f_0$ using $N_f=2+1+1$ lattice QCD. Finally, we compare both results for $\chi_\infty$, finding good agreement between quenched and dynamic computation.

\section{Lattice Actions}

\subsection{\texorpdfstring{$N_f=2+1+1$}{Nf=2+1+1} lattice QCD}

The calculation of the masses presented in this work was performed using configurations with $N_f=2+1+1$
dynamical Wilson twisted mass fermions at maximal twist generated by the European Twisted Mass Collaboration
(ETMC)~\cite{Baron:2010bv,Baron:2010th,Baron:2011sf}.

The lattice Wilson twisted mass fermion action\footnote{For a review on twisted mass fermions we refer to
e.g. \cite{Shindler:2007vp}} for the light sector
\cite{Frezzotti:2000nk,Frezzotti:2003ni,Frezzotti:2004wz}, i.e. $u$ and $d$ quarks, is given, in the
twisted mass basis, by:
\begin{equation}
 S_l[\psi,\psibar,U]=a^4 \sum_x \chibar_l(x)(D_W+m_0+i\mu_l\gamma_5\tau_3)\chi_l(x)\,,
\end{equation}
where $m_0$ and $\mu_l$ are the bare untwisted and twisted quark masses, respectively. 
$\chi_l=(\chi_u\, \chi_d)^T$ is a flavor doublet and $\tau_3$ acts in flavor space. 
The massless Wilson Dirac operator $D_W$ is defined as
\begin{equation}
 D_W=\frac{1}{2}(\gamma_\mu(\nabla_\mu+\nabla^\star)-a\nabla^\star_\mu\nabla_\mu)\,,
\end{equation}
where $\nabla_\mu$ and $\nabla_\mu^\star$ are the forward and backward covariant derivatives.

The lattice Wilson twisted mass action for the heavy doublet $\chi_h=(\chi_c\, \chi_s)^T$
\cite{Frezzotti:2004wz,Frezzotti:2003xj}, i.e. strange and charm quarks, is given by:
\begin{equation}
 S_h[\psi,\psibar,U]=a^4 \sum_x
\chibar_h(x)(D_W+m_0+i\mu_\sigma\gamma_5\tau_1+\mu_\delta\tau_3)\chi_h(x)\,.
\end{equation}
The bare twisted mass parameters $\mu_\delta$ and $\mu_\sigma$ are related to the bare strange and charm
quark masses through the following relation:
\begin{equation}
 \mu_{c,s}=\mu_\sigma\pm \frac{Z_P}{Z_S}\mu_\delta
 \label{eq:heavy_quark_masses}
\end{equation}
where $Z_P/Z_S$ defines the ratio of pseudoscalar and scalar flavor non-singlet renormalization
factors. 
Both doublets, $\chi_l$ and $\chi_h$, are related to their counterparts in the physical basis via chiral
rotations.

For the gauge sector the Iwasaki \cite{Iwasaki:1985we,Iwasaki:1996sn} gauge action was used which is
defined as:
\begin{equation}
  \label{eq:iwasaki}
  S_G[U]=\frac{\beta}{3}\sum_x\left(b_0\sum_{\genfrac{}{}{0pt}{}{\mu,\nu=1}{1\leq\mu<\nu}}^4\mathrm{Re}
\Tr \left(1-P^{1\times 1}_{x;\mu\nu}\right)+b_1\sum_{\genfrac{}{}{0pt}{}{\mu,\nu=1}{\mu \neq
\nu}}^4\mathrm{Re} \Tr \left(1-P^{1\times 2}_{x;\mu\nu}\right)\right)\,,
\end{equation}
where $b_1=-0.331$ and $b_0=1-8b_1$.
The dynamical ensembles used in this work are compiled in Tab.~\ref{tab:setup_nf211} with the labeling
adopted from Ref.~\cite{Baron:2010bv}.
For each $\beta$-value, there are several quark masses available, which allows one to study the chiral
extrapolation. 
The input parameters of our simulations are supplemented by Tab.~\ref{tab:tabNf211}, which contains the
chirally extrapolated values of the Sommer parameter $r_0/a$ for each value of $\beta$, which we use to
set the scale in our dynamical simulations. 
In addition, we quote in this table the chirally extrapolated values of the renormalization constant
ratio $Z\equiv Z_P/Z_S$ from the two definitions discussed in Ref.~\cite{Carrasco:2014cwa}. 
These two definitions differ by lattice artefacts and are, therefore, helpful in understanding the
corresponding systematic effects.

\begin{table}[t!]
 \centering
 \begin{tabular*}{1.\textwidth}{@{\extracolsep{\fill}}lcccccccc}
  \hline\hline
  ensemble & $\beta$ & $T/a \times (L/a)^3$ & $a\mu_\ell$ & $a\mu_\sigma$ & $a\mu_\delta$ & $N$ & $N_s$ &
$N_b$ \\ 
  \hline\hline
  $A30.32$   & $1.90$ & $64\times32^3$ & $0.0030$ & $0.150$  & $0.190$  & $1367$ & $24$ & $5$  \\
  $A40.24$   & $1.90$ & $48\times24^3$ & $0.0040$ & $0.150$  & $0.190$  & $2630$ & $32$ & $10$ \\
  $A40.32$   & $1.90$ & $64\times32^3$ & $0.0040$ & $0.150$  & $0.190$  & $863$  & $24$ & $4$  \\
  $A60.24$   & $1.90$ & $48\times24^3$ & $0.0060$ & $0.150$  & $0.190$  & $1251$ & $32$ & $5$  \\
  $A80.24$   & $1.90$ & $48\times24^3$ & $0.0080$ & $0.150$  & $0.190$  & $2449$ & $32$ & $10$ \\
  $A100.24$  & $1.90$ & $48\times24^3$ & $0.0100$ & $0.150$  & $0.190$  & $2493$ & $32$ & $10$ \\
  \hline
  $A80.24s$  & $1.90$ & $48\times24^3$ & $0.0080$ & $0.150$  & $0.197$  & $2517$ & $32$ & $10$ \\
  $A100.24s$ & $1.90$ & $48\times24^3$ & $0.0100$ & $0.150$  & $0.197$  & $2312$ & $32$ & $10$ \\
  \hline
  $B25.32$   & $1.95$ & $64\times32^3$ & $0.0025$ & $0.135$  & $0.170$  & $1484$ & $24$ & $5$  \\
  $B35.32$   & $1.95$ & $64\times32^3$ & $0.0035$ & $0.135$  & $0.170$  & $1251$ & $24$ & $5$  \\
  $B55.32$   & $1.95$ & $64\times32^3$ & $0.0055$ & $0.135$  & $0.170$  & $1545$ & $24$ & $5$  \\
  $B75.32$   & $1.95$ & $64\times32^3$ & $0.0075$ & $0.135$  & $0.170$  & $922$  & $24$ & $4$  \\
  $B85.24$   & $1.95$ & $48\times24^3$ & $0.0085$ & $0.135$  & $0.170$  & $573$  & $32$ & $2$  \\
  \hline
  $D15.48$   & $2.10$ & $96\times48^3$ & $0.0015$ & $0.120$  & $0.1385$ & $1045$ & $24$ & $10$ \\
  $D30.48$   & $2.10$ & $96\times48^3$ & $0.0030$ & $0.120$  & $0.1385$ & $474$  & $24$ & $3$  \\
  $D45.32sc$ & $2.10$ & $64\times32^3$ & $0.0045$ & $0.0937$ & $0.1077$ & $1887$ & $24$ & $10$ \\
  \hline\hline
 \end{tabular*}
 \caption{The dynamical $N_f=2+1+1$ simulations that are included in our investigations. The notation
that is used to label the ensembles is the same as in \cite{Baron:2010bv}. In addition to the value of
$\beta$, the lattice volumes and the bare quark masses $\mu_l$, $\mu_\sigma$, $\mu_\delta$ we give the
number of configurations $N$, the number of stochastic samples $N_s$ and the bootstrap block length $N_b$
that were used in the determination of flavor singlet
quantities~\cite{Ottnad:2012fv,Michael:2013vba,Michael:2013gka}.}
 \label{tab:setup_nf211}
\end{table}

\begin{table}[th!]
 \centering
 \begin{tabular}{@{\extracolsep{\fill}}ccccc}
  \hline\hline
  $\beta$ & $r_0/a$ & $a$ [fm] & $Z_P/Z_S$ (M1) & $Z_P/Z_S$ (M2) \\
  \hline\hline
  1.90 & 5.31(8) & 0.0885(36) & 0.699(13) & 0.651(6) \\
  1.95 & 5.77(6) & 0.0815(3)  & 0.697(7)  & 0.666(4) \\
  2.10 & 7.60(8) & 0.0619(18) & 0.740(5)  & 0.727(3) \\
  \hline\hline
 \end{tabular}
 \caption{Chirally extrapolated values of $r_0/a$ and $Z_P/Z_S$ from the two methods M1 and M2 as discussed in \cite{Carrasco:2014cwa} for our dynamical simulations.} 
 \label{tab:tabNf211}
\end{table}

\subsection{Quenched Action}

For the determination of $\chi_\infty$ in the pure YM theory, we have generated four quenched
ensembles at four different values of the lattice spacing. 
For the gauge action, we used again the Iwasaki action -- Eq.~(\ref{eq:iwasaki}).

We emphasize that, in order to test the Witten-Veneziano formula, we tried to achieve a very
similar setup for the quenched and dynamical situation -- we took the same gauge action, we matched the
physical volume and took a fixed value of $r_0\mu$ in the valence Dirac operator used for spectral
projectors, equal to $r_0\mu$ of the dynamical simulations\footnote{We do not have available the value
of the renormalization constant $Z_P$ for the quenched simulations, so this implies only approximate
matching of valence quark masses, assuming that with the same gauge action $Z_P$ is not very different.}.
The details of determining $\kappa_c$ needed for $\mathcal{O}(a)$ improvement can be found in
\ref{sec:appendixA}. 
All the details of the quenched simulations are summarized in Tab.~\ref{tab:qchddetails}.

The quenched simulations have been performed using the HMC algorithm implemented in the tmLQCD  package \cite{Jansen:2009xp}.
The usage of this algorithm might introduce somewhat larger autocorrelation times compared to e.g. the
heatbath with overrelaxation algorithm. However, compared to the dynamical simulations, the generation of quenched ensembles was still only a
small effort.

\begin{table}[t!]
 \centering
 \begin{tabular}{@{\extracolsep{\fill}}ccccccc}
  \hline\hline
  $\beta$ & $T/a \times (L/a)^3$ & $r_0/a$ & $a$ [fm] & $a\mu$ & $r_0\mu$& $\kappa_c^{\chi}$  \\
  \hline\hline
  2.37 & $40\times20^3$ & 3.59(2)(3) & 0.1393(14) & 0.0087 & 0.0312(3) & 0.158738 \\
  2.48 & $48\times24^3$ & 4.28(1)(5) & 0.1182(14) & 0.0073 & 0.0309(4) & 0.154928 \\
  2.67 & $64\times32^3$ & 5.69(2)(3) & 0.0879(6)  & 0.0055 & 0.0314(2) & 0.150269 \\
  2.85 & $80\times40^3$ & 7.29(7)(1) & 0.0686(7)  & 0.0043 & 0.0313(3) & 0.147180 \\
 \hline\hline
 \end{tabular}
 \caption{All relevant parameters of the pure gauge ensembles for $\beta=2.37,\; 2.48, \; 2.67$ and $2.85$. The errors of $r_0/a$ correspond to statistical and systematic uncertainties, respectively.}
 \label{tab:qchddetails}
\end{table}

\section{Topological susceptibility in the pure Yang-Mills theory}

\subsection{The spectral projectors method}

In order to compute the topological susceptibility ($\chi_{top}$) in the pure YM theory (for this case,
we will denote it specifically by $\chi_\infty$), we used the method of spectral projectors
\cite{LuscherGiusti}. 
This method was originally applied to the pure gauge theory in
Ref.~\cite{Luscher:2010ik}
and then to the dynamical case for twisted mass fermions in
Ref.~\cite{Cichy:2013rra}.

We introduce the definition of the topological susceptibility in terms of the
spectral projector $\RM$ and refer to the original papers for
further details about the method \cite{LuscherGiusti,Luscher:2010ik}:
\begin{equation}
 \label{eq:topsusspproj}
 \chitop=\frac{Z_S^2}{Z_P^2}\frac{1}{V}\expect{\Tr\{\gamma_5 \RM^2\}\Tr\{\gamma_5 \RM^2\}}\,,
\end{equation}
where the ratio of renormalization constants $Z_P/Z_S$ differs from unity due to the use of
non-chirally symmetric Wilson-type fermions. 
The spectral projector $\RM$ is an orthogonal projector to the subspace of fermion fields spanned by the
eigenvectors of the massive Hermitian Dirac operator $D^\dagger D$ (with $D=D_W+m_0+i\mu_l\gamma_5\tau^3$
in our case) with eigenvalues not larger than $M^2$.
To achieve $\mathcal{O}(a^2)$ scaling towards the continuum limit, the renormalized value of the
threshold $M$, denoted by $M_R$, has to be fixed for all ensembles.
The value of $M_R$ can be chosen arbitrarily, but it is advisable to avoid too small values (close to
the renormalized quark mass) and choose $aM_R\ll1$ to avoid enhanced cut-off effects
\cite{LuscherGiusti}.

In practice, we compute the following spectral observables:
\begin{align}
 & \A = \frac{1}{N}\sum_{k=1}^N(\RM^2\eta_k,\;\RM^2\eta_k)\,, \label{eq:obsA}\\
 & \B = \frac{1}{N}\sum_{k=1}^N(\RM\gamma_5\RM\eta_k,\;\RM\gamma_5\RM\eta_k)\,, \label{eq:obsB}\\
 & \C = \frac{1}{N}\sum_{k=1}^N(\RM\eta_k,\;\gamma_5\RM\eta_k)\,, \label{eq:obsC}
\end{align}
where $N$ is the number of stochastic sources $\eta_k$ ($k=1,\ldots,N$) used for the construction of the
spectral projector, solving the Dirac equation $(D^\dagger D+M^2)\psi=\eta_k$ an appropriate number of
times.
The above observables are directly related to the right hand side of Eq.~(\ref{eq:topsusspproj}) through
the following expression:
\begin{equation}
 \label{eq:C2BN}
 \expect{\Tr\{\gamma_5 \RM^2\}\Tr\{\gamma_5 \RM^2\}}=\expect{\C^2}-\frac{\expect{\B}}{N}=\expect{\C^{'2}}\,.
\end{equation}
Since the use of two independent sets of sources to compute the square of $\C$ would duplicate the cost
of the calculation, we need to introduce the correction given by $\expect{\B}/N$ to correct for the bias
introduced by the usage of the same set of stochastic sources to compute the square of $\C$.

In order to compute the ratio of renormalization constants $Z_P/Z_S$, we also use the spectral
projector method, since the low cost of the calculation allows us to obtain reliable
estimates of $Z_P/Z_S$ not available a priori, since the quenched ensembles were generated specifically
for this project. The ratio can be obtained through the following expression:
\begin{equation}
 \label{eq:ZPoverZSeq}
 \frac{Z_P^2}{Z_S^2}=\frac{\expect{\A}}{\expect{\B}} \,.
\end{equation}
This application was first proposed in Ref.~\cite{LuscherGiusti} and applied for dynamical simulations
of twisted mass fermions in Ref.~\cite{Cichy:2013gja}. In the latter reference, a comparison of the
result given by spectral projectors and more standard methods like RI-MOM \cite{Alexandrou:2012mt} and
$x$-space \cite{Cichy:2012is} was done and compatible results were found. 
A study of discretization effects and finite volume effects was also presented.

\subsection{Computation of \texorpdfstring{$Z_P/Z_S$}{ZP over ZS}}

As we mentioned in previous section, in order to compute the topological susceptibility using twisted
mass fermions, the final value has to be renormalized using the ratio $Z_P/Z_S$, which is only equal to
unity in a chirally symmetric theory.

\begin{figure}[!h]
 \begin{center}
  \includegraphics{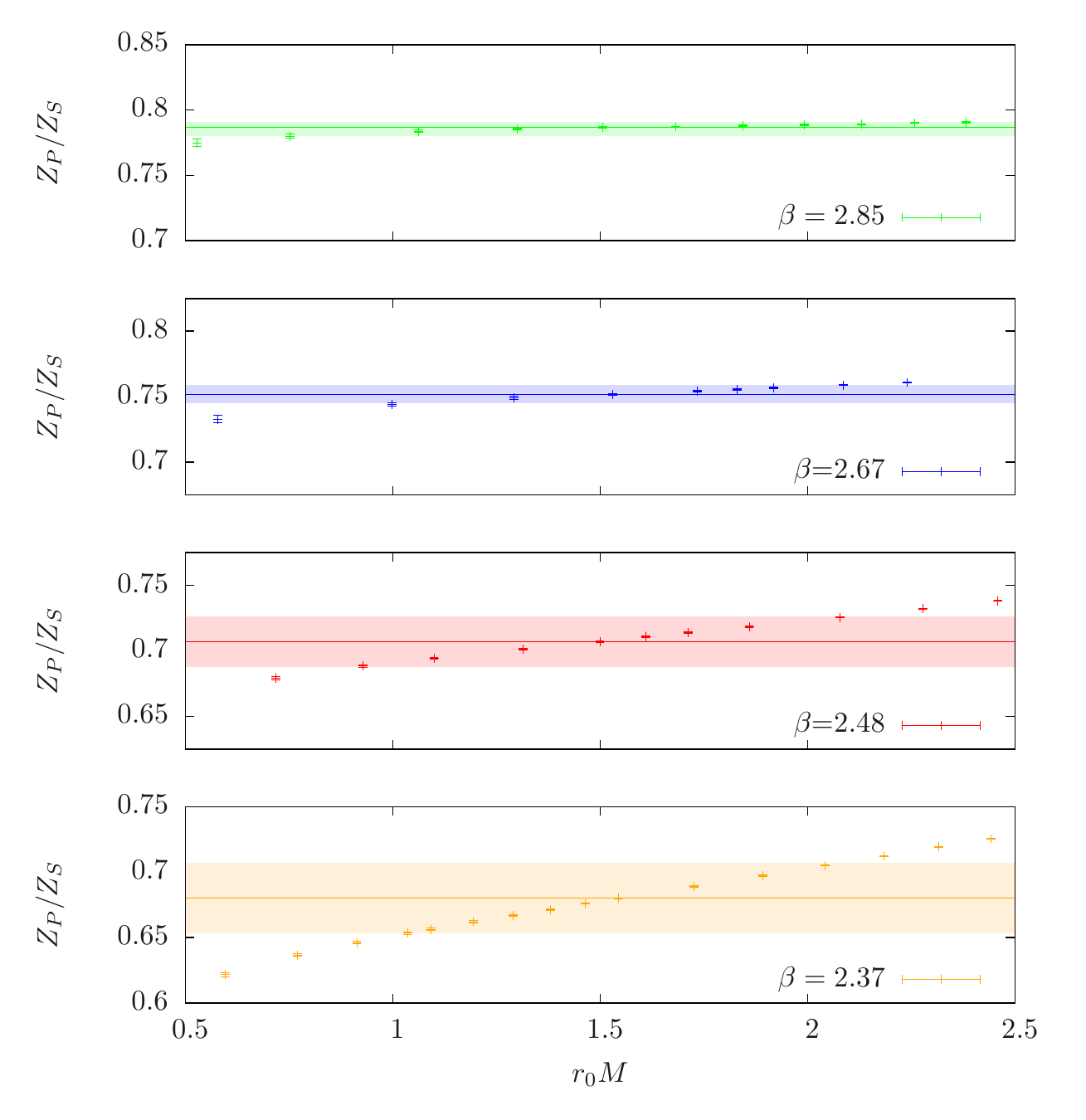}
 \end{center}
 \caption{\label{fig:qchdzpzsfnal} Results of $Z_P/Z_S$ as a function of $r_0M$ for quenched ensembles. The straight lines correspond to the final result and the shaded areas to the systematic error. For further details we refer to the text.}
\end{figure}

\begin{table}[!h]
 \centering
 \begin{tabular}{@{\extracolsep{\fill}}cc}
  \hline\hline
   $\beta$ &  $Z_P/Z_S$ \\
  \hline\hline
  2.37  & 0.680(1)(27)\\
  2.48  & 0.707(1)(19)\\
  2.67  & 0.752(1)(7)\\
  2.85  & 0.787(1)(3)\\
  \hline\hline
 \end{tabular}
 \caption{Results for $Z_P/Z_S$ values using spectral projectors for the quenched ensembles. 
   The errors quoted correspond to, first, statistical and, second, systematic uncertainty (coming from
the dependence on $M$).}
 \label{tab:qchdzpzstab}
\end{table}

In Fig.~\ref{fig:qchdzpzsfnal}, the value of $Z_P/Z_S$ for different values of $r_0M$ for
the four quenched ensembles listed in Tab.~\ref{tab:qchddetails} is shown. 
We follow the strategy presented in Ref.~\cite{Cichy:2013rra} to extract the value of
$Z_P/Z_S$ for each ensemble, i.e. we take the central value to be the
one corresponding to some large value of $r_0 M$, chosen as 1.5, in
order to avoid contamination by non-perturbative effects appearing for
small $r_0 M$. However, due to cut-off effects, even at large $r_0 M$,
we observe some residual dependence on $r_0 M$. To quantify the
systematic error coming from this dependence, we consider the spread
of $Z_P/Z_S$ in the interval $r_0 M\in[1,2]$, i.e. this systematic
uncertainty is taken as the larger of the deviations of $Z_P/Z_S(r_0
M=1.5)$ with respect to $Z_P/Z_S(r_0 M=1)$ and $Z_P/Z_S(r_0 M=2)$. As
expected, this uncertainty drops down significantly with decreasing
lattice spacing. Thus, we obtain the results presented in Tab.~\ref{tab:qchdzpzstab}. 
The first error corresponds to the statistical error given by the 
spectral projectors method, where autocorrelations were taken into account. 
The second error corresponds to the systematic error introduced in our 
calculation through the residual dependence of $Z_P/Z_S$ on $M$.

\subsection{Continuum limit of \texorpdfstring{$\chi_{\infty}$}{quenched chi}}

The main aim of this work is to test the Witten-Veneziano formula. To this end, we need to compute a
reliable continuum limit of the topological susceptibility in the quenched case. The physical conditions,
such as the volume or the action used, were matched to the situation of the dynamical simulations
used to compute the masses. 

Following this strategy, we generated the first ensemble using the Iwasaki gauge action at $\beta=2.67$
with a value of $r_0/a$ matching the one of the dynamical ensemble at $\beta=1.95$. The other
characteristics, such as the volume was matched to the ensemble B55.32 (see
table~\ref{tab:setup_nf211}) with a $32^3\times 64$ lattice and the value of the valence quark mass for
the spectral projector method was set to $a\mu=0.0055$.

To obtain the continuum limit we needed to extend the simulations to different lattice
spacings. Thus, we kept the physical volume constant by imposing a constant the physical extent of the
lattice. In particular, if we assume $r_0=0.5~$fm, the physical volume was kept at $L\approx2.8~$fm.
Similarly, for the quark mass, we demanded the product $r_0\mu$ to remain invariant. As already pointed
out in the previous section, we do not have any estimate of $Z_P$ and hence we can only match the bare
product $r_0\mu$, assuming that the changes in $Z_P$ are small. All the exact values
for each ensemble can be found in Tab.~\ref{tab:qchddetails}. 

As we mentioned in the introduction, the spectral projector method requires a mass input parameter $M$.
In principle, the continuum limit result should be independent of this value, as long as it is kept fixed
for all the ensembles entering the calculation. As the value of the renormalization constant $Z_P$
needed to  renormalize the input parameter $M$ is not available, an alternative strategy
was followed to match the values of $M_R$. We used the fact that the mode number remains constant
at a fixed value of $M_R$ and in a constant physical volume \cite{LuscherGiusti}, i.e.
\begin{equation}
 \frac{a_1}{a_2}=\left(\frac{\nu_1 n_2}{\nu_2 n_1}\right),\,
\end{equation}
where $a_i$ is the lattice spacing and $n_i$ the number of lattice points of the ensembles. 
Hence, the condition
In Tab.~\ref{tab:tab1chiinfty}, the values of the mode number given by the spectral observable $\A$ are
shown.

\begin{table}[!t]
 \centering
 \begin{tabular}{@{\extracolsep{\fill}}ccccccc}
  \hline\hline
  $\beta$ & $\expect{\A}$ & $\tau_{\rm  int}$ & $\expect{\B}$ & $\tau_{\rm int} $& $\expect{\C}$ & $\tau_{\rm int}$\\
  \hline\hline
  2.37 & 79.4(2) & 0.5(1) & 17.51(5) & 0.5(1) &  0.19(19) & 0.5(1) \\
  2.48 & 78.7(2) & 0.5(1) & 22.00(8) & 0.5(1) &  0.29(32) & 0.4(1) \\
  2.67 & 78.5(3) & 0.5(1) & 29.4(2)  & 0.5(1) & -0.61(64) & 0.7(2) \\
  2.85 & 78.1(4) & 0.5(1) & 36.5(2)  & 0.4(1) &  0.93(93) & 0.8(3) \\
  \hline\hline
 \end{tabular}
 \caption{Results of $\expect{\A},\;\expect{\B}\;{\rm and} \;\expect{\C} $ their corresponding values of
$\tau_{\rm int}$ for quenched  ensembles at four different lattice spacings. Errors are statistical
only. $\tau_{\rm int}$ is in the units of the taken step between measured configurations, which was
increasing for increasing $\beta$ (see also Tab.~\ref{tab:tab2chiinfty}).}
 \label{tab:tab1chiinfty}
\end{table}

Once we know the value of the ratio $Z_P/Z_S$ for each $\beta$ value, we are prepared to
compute the topological susceptibility $\chi_{\rm top}$.  

The continuum limit is a crucial aspect of lattice QCD. In fact, the rate at which a quantity approaches
the continuum limit plays a fundamental role in lattice calculations, since it is directly related to the
accuracy of the final computation of physical observables.

The twisted mass action at maximal twist guarantees the $\obs(a)$ improvement for on shell observables
\cite{Frezzotti:2003ni}. The problem can arise if our observable is affected by short distance
singularities, since it can spoil the, otherwise guaranteed, $\obs(a^2)$ scaling. 

\begin{figure}[!t]
 \begin{center}
  \includegraphics[width=1.0\textwidth]{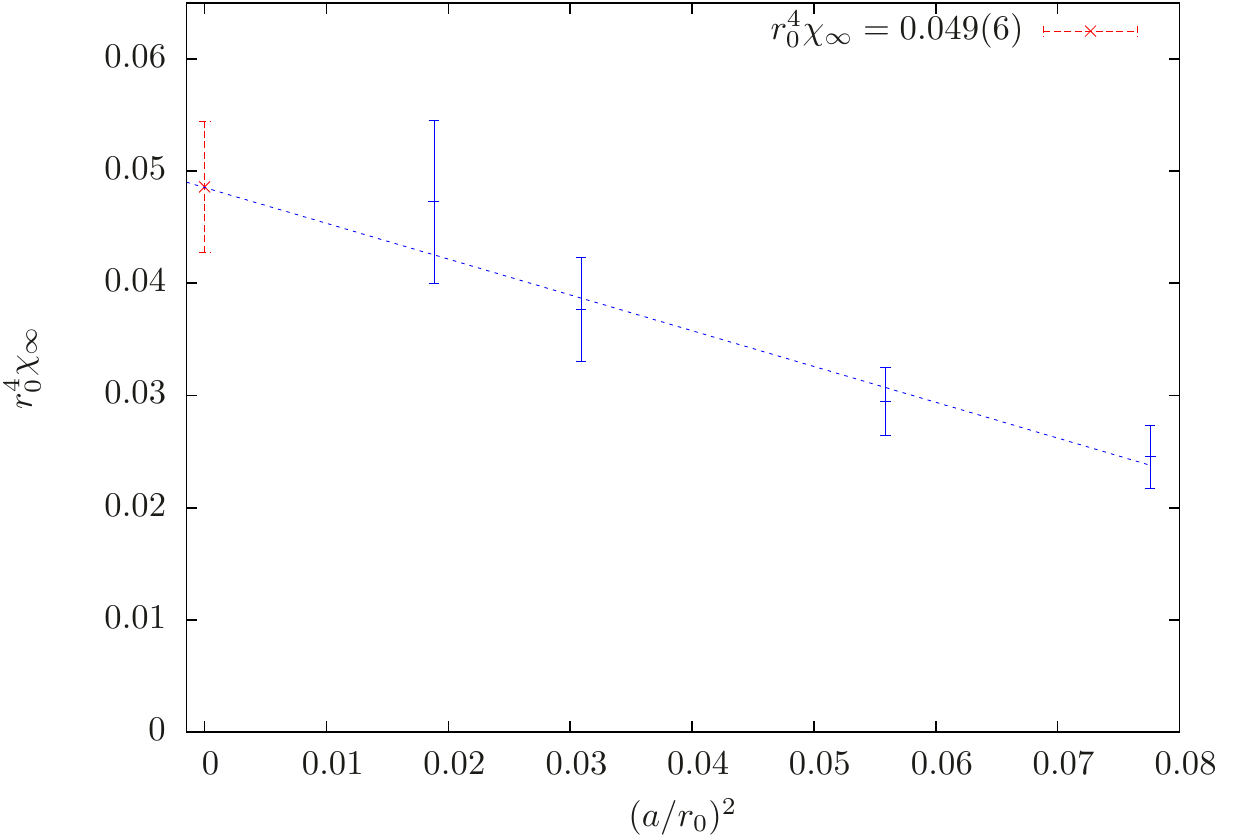}
 \end{center}
 \caption{\label{fig:clim_ts}Continuum limit extrapolation of $\chi_\infty$ as a function of $(a/r_0)^2$ for our quenched ensembles.}
\end{figure}

\begin{table}[!t]
 \centering
 \begin{tabular}{@{\extracolsep{\fill}}cccccc}
  \hline\hline
  $\beta$ & $N_s$ & $N_{\rm traj}$ & $N_{\rm meas}$  & $\expect{\C'^2}$& $r_0^4\chi_{\rm top}$\\
  \hline\hline
  2.37 & 6 &  78000 & 769 & 25.9(1.3) & 0.0250(14)(22)(8)  \\
  2.48 & 6 &  98000 & 412 & 32.8(2.4) & 0.0295(24)(16)(8) \\
  2.67 & 6 & 132000 & 157 & 47.5(5.1) & 0.0376(45)(7)(8)  \\
  2.85 & 6 & 380000 & 119 & 60.9(7.9) & 0.0485(70)(4)(18) \\
  \hline\hline
 \end{tabular}
 \caption{Results of $\chi_{\rm top}$ for the continuum limit in the pure gauge theory. $N_s$ represents
the number of stochastic sources whereas $N_{\rm meas}$ is the number of evaluated
independent configurations and $N_{\rm traj}$ gives the corresponding number of thermalized MC
trajectories. The errors quoted for $\chi_{\rm top}$ are statistical, coming from $Z_P/Z_S$ and from
$r_0/a$, respectively.}
 \label{tab:tab2chiinfty}
\end{table}

Our observable, as discussed in detail in Refs.~\cite{Cichy:2014yca,Cichy:2013zea}, is affected by short
distance singularities. However, all terms linear in the lattice spacing that are consequently arising in the Symanzik expansion vanish at
maximal twist \cite{Cichy:2014yca}. Thus, our observable remains $\obs(a)$ improved, guaranteeing an $\obs(a^2)$
scaling towards the continuum limit.

Consequently, we have performed a linear extrapolation in $(a/r_0)^2$ to extract the continuum limit
which leads to the following continuum result: 
\begin{equation}
 r_0^4\chi_\infty= 0.049(6)_{\rm stat+sys}\,,
\end{equation}
where the error is dominated by statistical uncertainty, but takes into account also the systematic ones (combined in quadrature with the statistical errors).
In Fig.~\ref{fig:clim_ts}, the results of the topological susceptibility for different values of the
lattice spacing together with the continuum limit extrapolation are plotted. All the intermediate and
final results are compiled in Tab.~\ref{tab:tab1chiinfty} and Tab.~\ref{tab:tab2chiinfty}, respectively,
as well as other relevant details.

The continuum value for the topological susceptibility can be compared to earlier results using chiral invariant lattice fermions \cite{DelDebbio:2004ns}, $r_04\chi_\infty=0.059(3)_{\rm stat}$ and
from spectral projector methods which avoid short distance singularities \cite{Luscher:2010ik}, $r_0i^4\chi_\infty=0.061(6)$. We note that our result is in agreement with the approach used in \cite{Durr:2006ky},
$r_0^4\chi_\infty=0.0524(7)_{\rm stat}(6)_{\rm sys}$, which has been obtained from a combined continuum and infinite volume limit. For the older result in \cite{DelDebbio:2004ns} only statistical
errors have been quoted and finite size effects could not be resolved within the statistical accuracy. However, the lattices in the present study exhibit a significantly larger volume (the physical value of $L$ is
about a factor two larger), while the range of lattice spacings is similar which might explain the difference in the final values for $r_0^4\chi_\infty$. In general, older studies \cite{Lucini:2001ej,DelDebbio:2002xa,DelDebbio:2003rn,Giusti:2003gf} tend
to favor larger values for the topological susceptibility as has already been remarked in \cite{Durr:2006ky}.

\subsection{Autocorrelations}

We close this section by a discussion of autocorrelation effects in the quenched simulations, which
might affect significantly the errors.

The topological charge is a quantity highly affected by autocorrelations when the continuum limit is
approached due to appearance of topological barriers, i.e. the transitions between topological sectors
are suppressed.  For this reason, particularly long Monte Carlo simulations are needed in order to
guarantee that all topological sectors are sampled adequately.  

\begin{figure}[t]
 \begin{center}
  \includegraphics[width=0.49\textwidth]{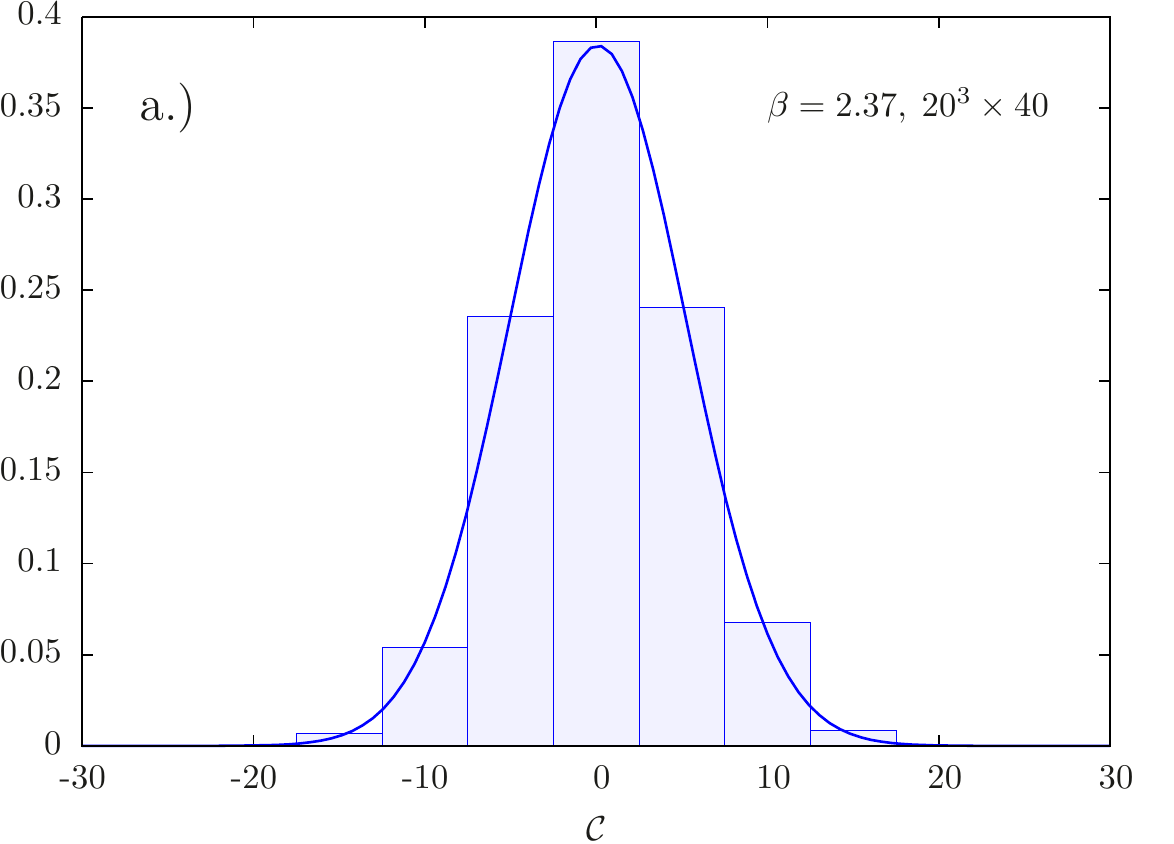}
  \includegraphics[width=0.49\textwidth]{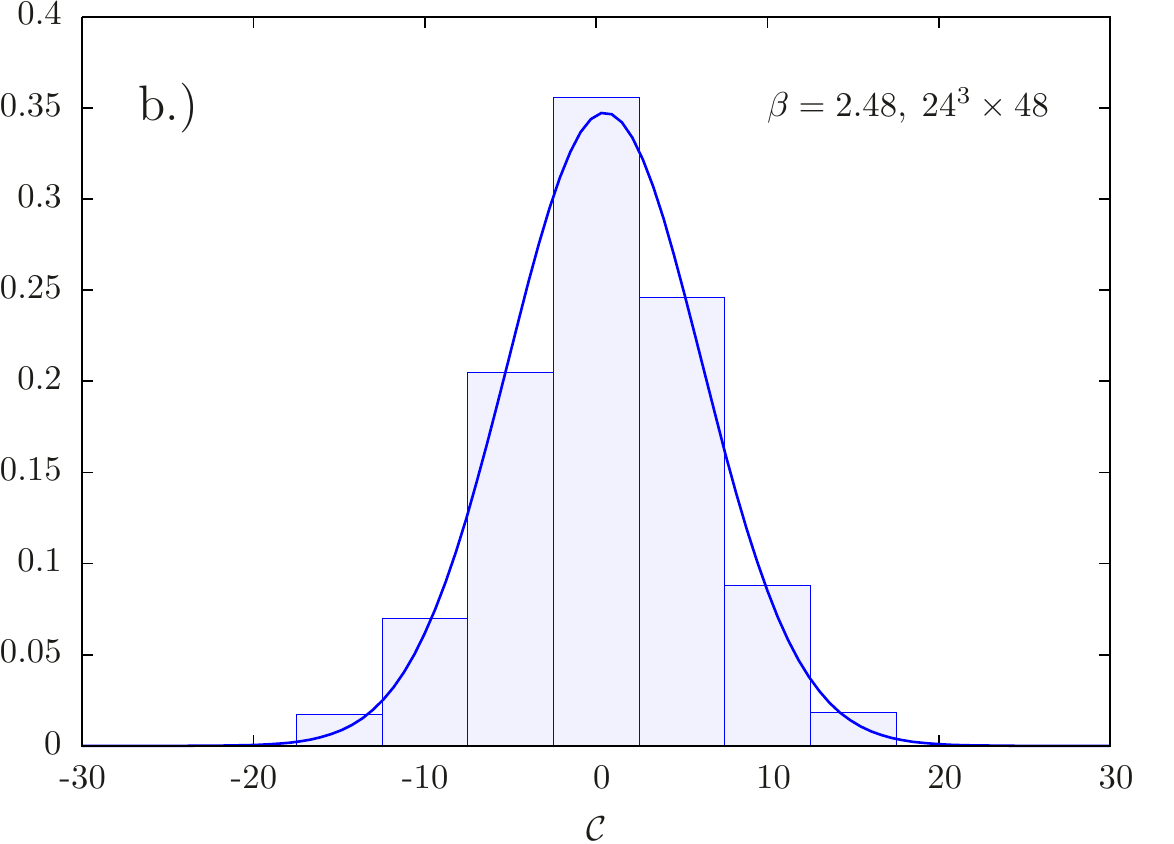}\\
  \includegraphics[width=0.49\textwidth]{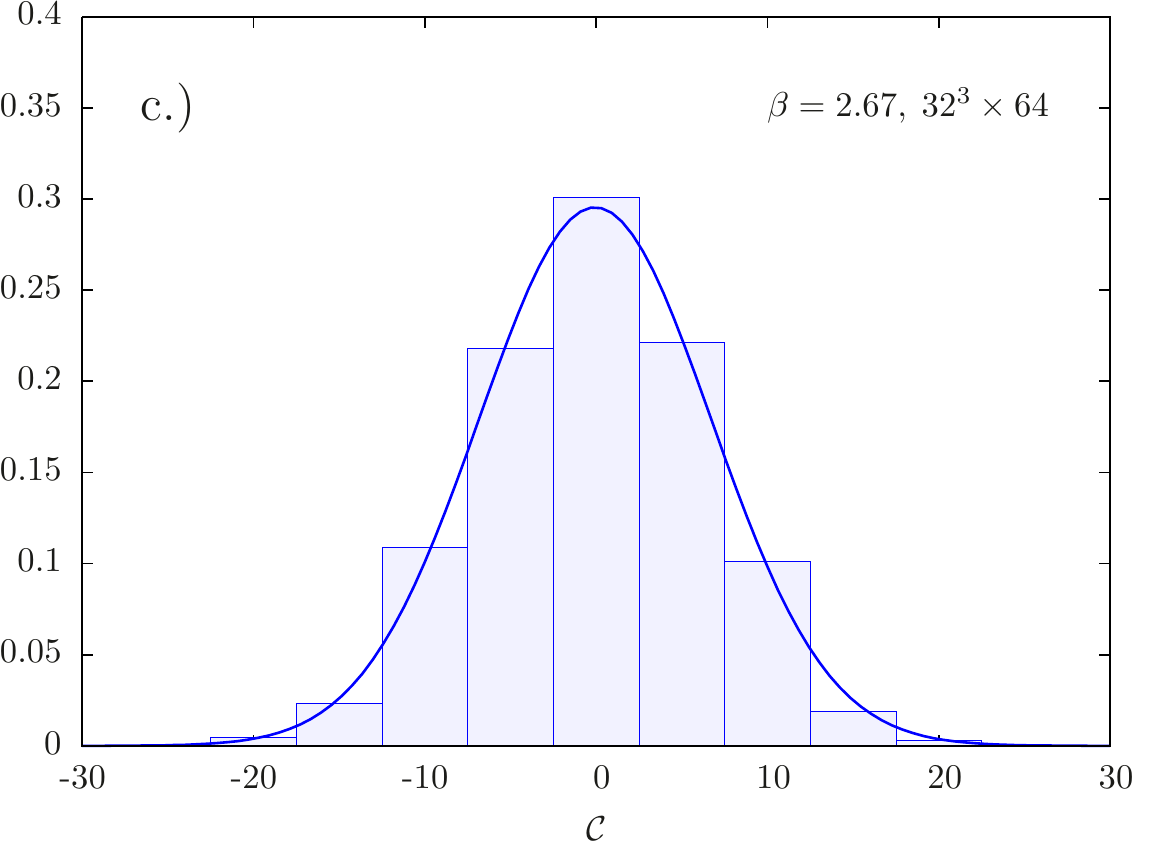}
  \includegraphics[width=0.49\textwidth]{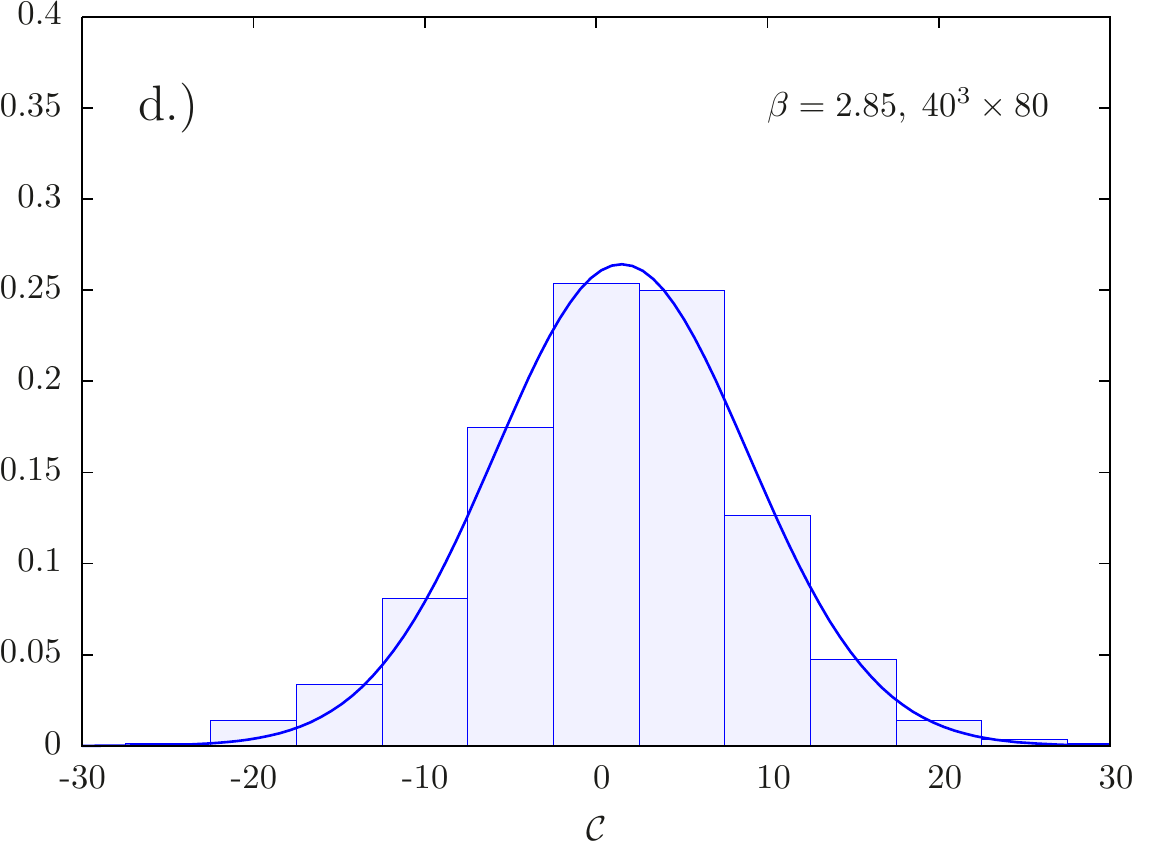}
 \end{center}
 \caption{\label{fig:hist_Cquenched} Histograms of the observable $\C$ for four quenched ensembles at $\beta=2.37$ (a.), $\beta=2.48$ (b.), $\beta=2.67$ (c.) and $\beta=2.85$ (d.).}
\end{figure}

The topological charge is expected to follow a Gaussian distribution centered at zero in the large volume
regime. Consequently, due to the fact that  the stochastic observable $\C$ is closely related to the
topological charge $Q$,  we expect the same behavior for the distribution of $\C$ and, equally,
$\expect{C}=0$.  

In Fig.~\ref{fig:hist_Cquenched}, the histograms of the stochastic observable $\C$ are  shown for all the
quenched  ensembles at four different values of the lattice spacing. In all cases we were able to
construct a histogram compatible with a Gaussian distribution within errors.

\begin{figure}[t]
 \begin{center}
  \includegraphics[width=0.49\textwidth]{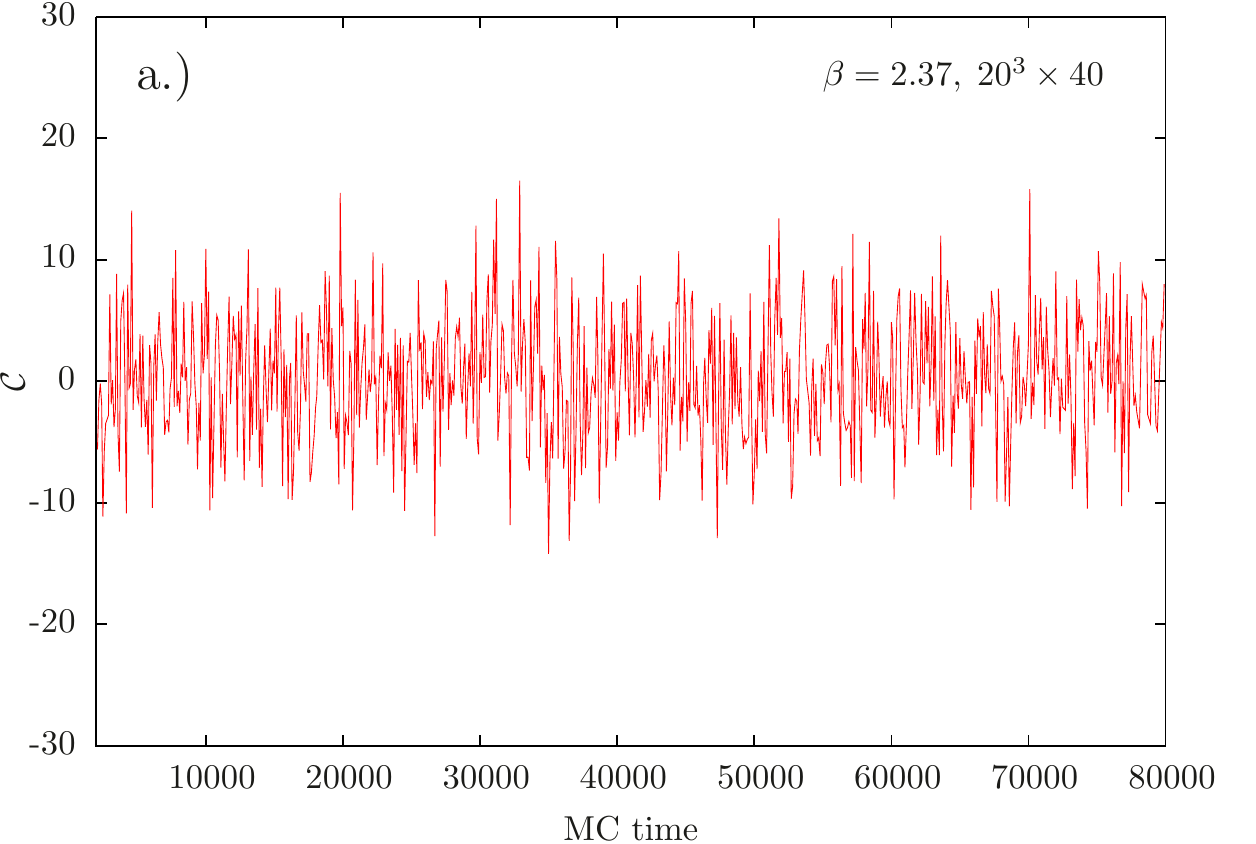}
  \includegraphics[width=0.49\textwidth]{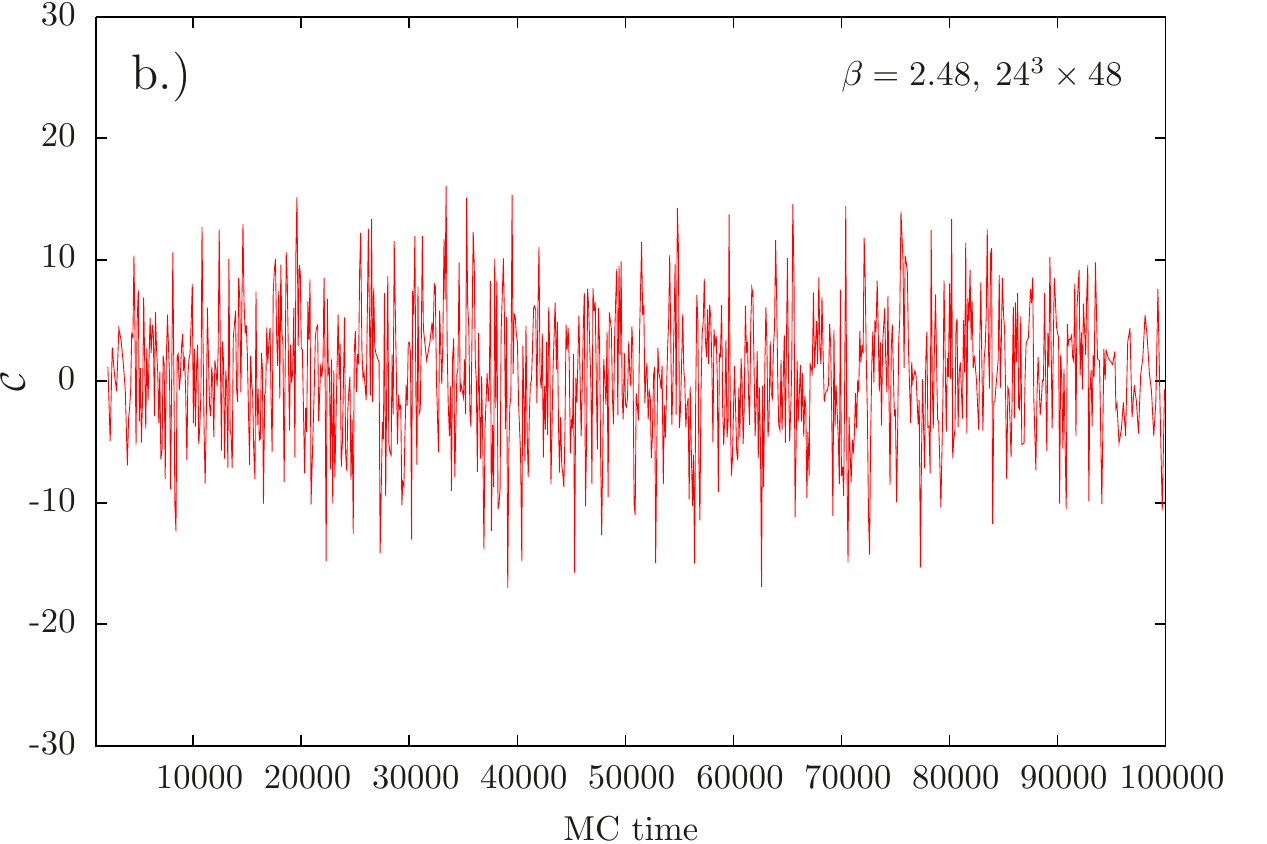}\\
  \includegraphics[width=0.49\textwidth]{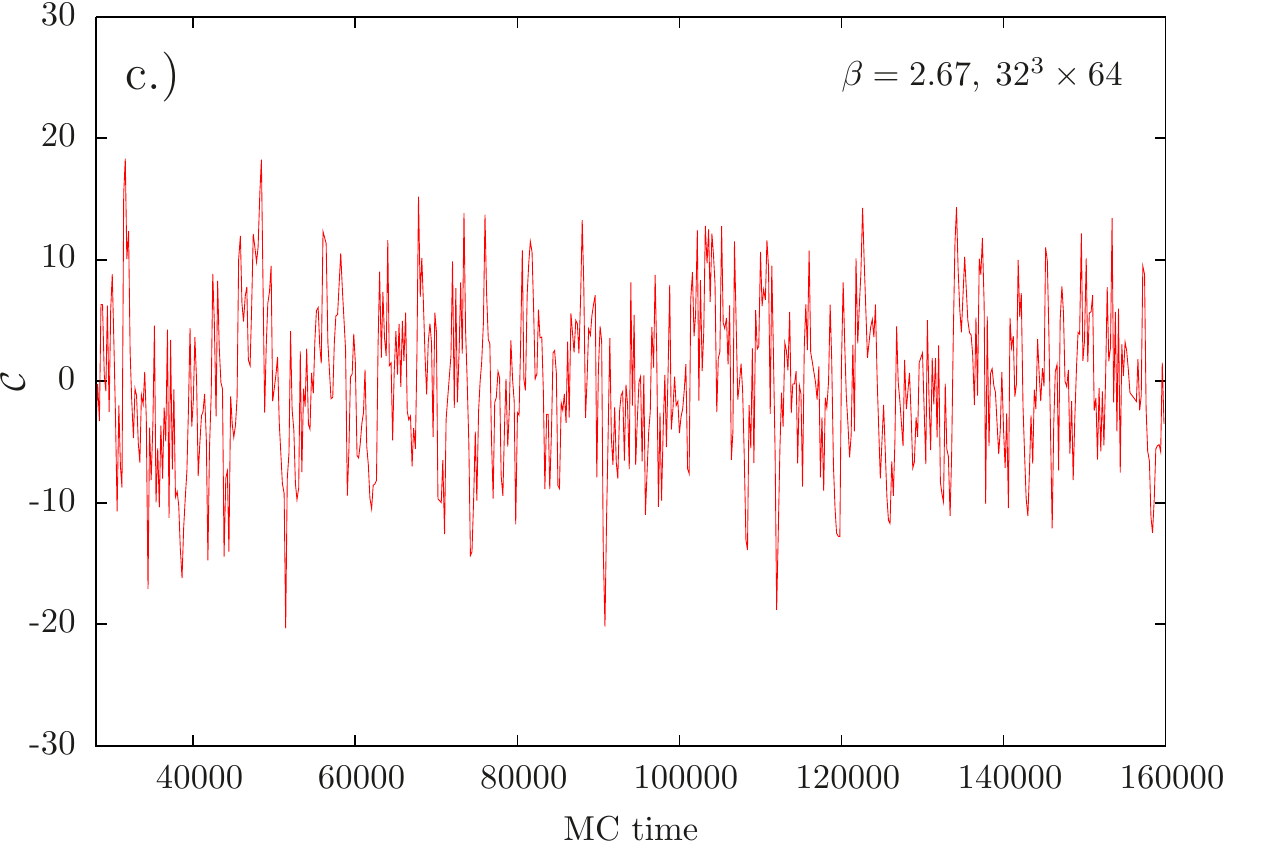}
  \includegraphics[width=0.49\textwidth]{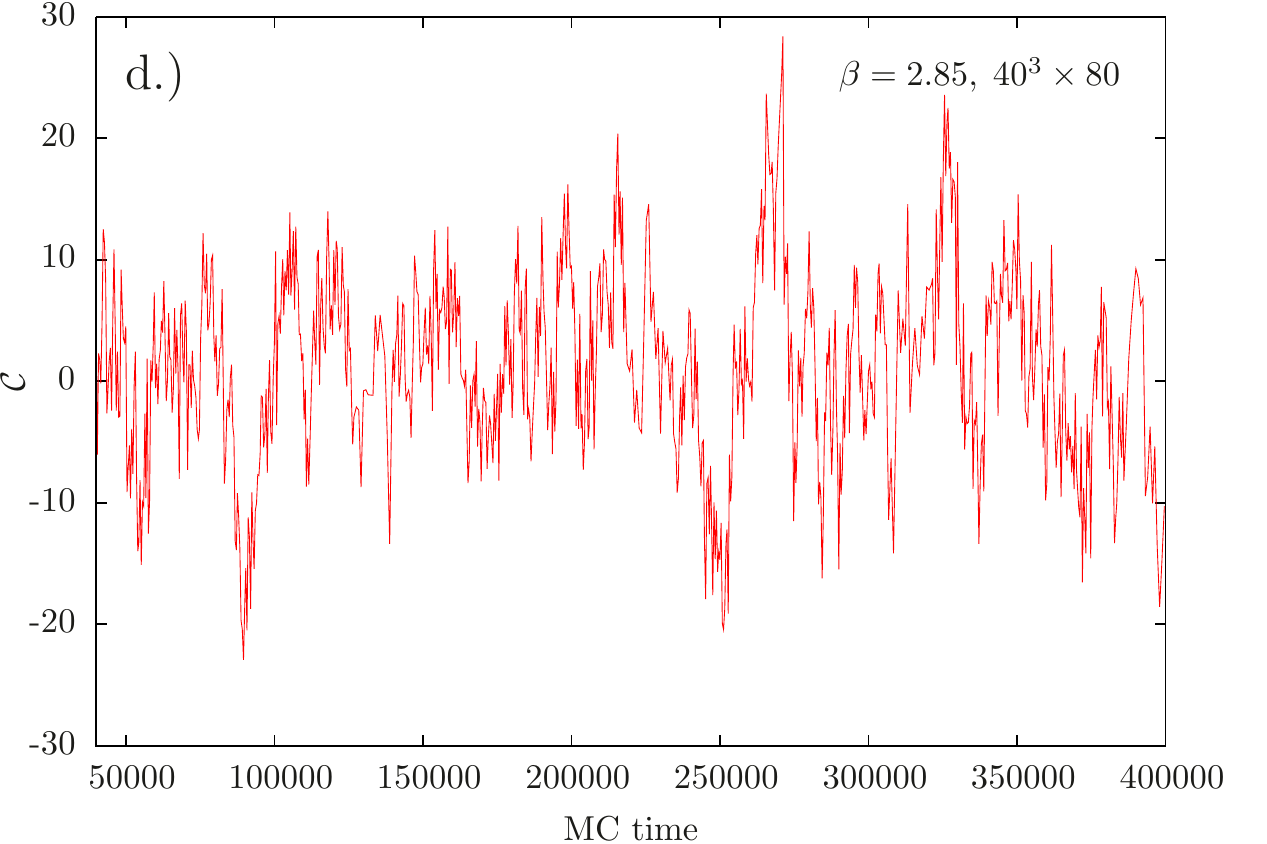}
 \end{center}
 \caption{\label{fig:MC_Cquenched} Monte Carlo histories of the observable $\C$ for four quenched ensembles at $\beta=2.37$ (a.), $\beta=2.48$ (b.), $\beta=2.67$ (c.) and $\beta=2.85$ (d.). }
\end{figure}

Fig.~\ref{fig:MC_Cquenched} shows the Monte Carlo history of the observable $\C$ again for all the
ensembles. 
It is clear, even from visual inspection of the plots, that autocorrelations become significant for the
finest lattice spacing. Therefore, even though we generated a few times more trajectories for this
ensemble than for the other ones, we only obtained an estimate of the topological susceptibility with a much
higher statistical error, since we had to take a step of around 3200 trajectories between measurements to
obtain $\tau_{\rm int}$ compatible with 0.5, i.e. we had fewer than 120 independent gauge field
configurations (see also Tab.~\ref{tab:tab2chiinfty}).

\section{Fermionic contributions to the Witten-Veneziano formula from dynamical simulations}
Having established a reliable way to extract $\chi_\infty$ from the pure Yang-Mills theory, we still need
to tackle the l.h.s. of Eq.~(\ref{eq:W-Vformula}). In this section, we discuss how to compute the
relevant masses and the flavor singlet decay constant parameter $f_0$. The computations in the flavor
singlet sector are very demanding, because of large contributions from quark disconnected diagrams.
Moreover, the computation of $f_0$ on the lattice turns out to be technically involved, due to the fact
that the axial vector matrix elements cannot be computed with sufficient signal-to-noise ratio.
Therefore, we need to consider pseudoscalar matrix elements and resort to $\chi$PT to relate them to the
desired axial vector ones. Further complications arise from the fact that the interpolating operators on
the lattice are defined in the so-called quark flavor basis, whereas decay constants and the
corresponding matrix elements are commonly defined in the singlet-octet flavor basis. This issue can be
circumvented in the framework of $\chi$PT as well.

\subsection{Computations in the flavor singlet sector}
The masses for $\eta$ and $\eta'$ mesons are computed as described in
\cite{Ottnad:2012fv,Michael:2013gka}. In particular, we employed the method of subtracting excited states
in the quark connected part of the correlation functions \cite{Neff:2001zr}. Together with the
application of the one-end trick for the evaluation of quark disconnected diagrams in the light quark
sector \cite{Boucaud:2008xu}, this yields a substantial improvement of the resulting statistical errors
\cite{Jansen:2008wv,Michael:2013gka}. Here, we will only outline some basic details that are relevant for
the discussion of the flavor singlet decay constant parameter $f_0$ in the next section. We use
pseudoscalar operators in the physical basis:
\begin{align}
 \mathcal{P}_l^{0,phys} &= \frac{1}{\sqrt{2}} \bar{\psi}_l i \g{5} \psi_l \,, \\
 \mathcal{P}^{\pm,phys}_h &= \bar{\psi}_h i \g{5} \frac{1\pm\tau^3}{2}\psi_h \,,
\end{align}
where $\bar{\psi}_l$, $\psi_l$ and $\bar{\psi}_h$, $\psi_h$ refer to degenerate light and non-degenerate
heavy quark doublets, respectively. The doublet structure and the flavor projector $(1\pm\tau^3)/2$ is
required due to the twisted mass formulation and the flavor projector in the second lines allows to
consider the non-degenerate charm and strange components separately. At maximal twist the corresponding
operators read
\begin{align}
 \mathcal{P}_l^{0,phys} &\rightarrow -\frac{1}{\sqrt{2}}\bar{\chi}_l \tau^3 \chi_l & \equiv \mathcal{S}_l^{3,tm} \,, \label{eq:light_S_eta_operator}\\
 \mathcal{P}^{\pm,phys}_h &\rightarrow \frac{1}{2}\bar{\chi}_h \l(-\tau^1 \pm i \g{5} \tau^3 \r) \chi_h & \equiv \mathcal{P}_h^{\pm,tm} \,, \label{eq:heavy_P_tm_eta_operators}
\end{align}
with $\mathcal{S}$ the scalar density.
In the following, we will neglect the charm operator, as it does neither contribute to the $\eta$ nor the
$\eta'$ within errors, i.e. we consider only the strange component $\mathcal{P}_h^{-,tm} \equiv
\mathcal{P}_s^{tm}$. The mixing between the flavor non-singlet scalar and pseudoscalar currents in the
heavy sector of the twisted basis introduces a relative factor of $Z_S/Z_P$ under renormalization:
\begin{equation}
 \mathcal{P}_s^{tm,r} = \frac{1}{2} Z_P \bar{\chi}_h \l(-\frac{Z_S}{Z_P}\tau^1 - i \g{5} \tau^3\r) \chi_h \,. \label{eq:heavy_P_tm_eta_operator_renormalized}
\end{equation}
Similarly, we apply a factor $Z_S/Z_P$ instead of $Z_S$ to the light operator $\mathcal{S}_l^{3,tm}$
allowing to pull out a global factor of $Z_P^2$ from the resulting, renormalized correlation function
matrix:
\begin{equation}
 \mathcal{C}^r(t) = Z_P^2 \tilde{\mathcal{C}}(t) \,,
\end{equation}
where
\begin{equation}
 \tilde{\mathcal{C}}(t) = \l(\begin{array}{cc}
<\tilde{\mathcal{S}}_l^{3,tm}(t)\tilde{\mathcal{S}}_l^{3,tm}(0)> &
<\tilde{\mathcal{S}}_l^{3,tm}(t)\tilde{\mathcal{P}}_s^{tm}(0)> \\
<\tilde{\mathcal{P}}_s^{tm}(t)\tilde{\mathcal{S}}_l^{3,tm}(0)> &
<\tilde{\mathcal{P}}_s^{tm}(t)\tilde{\mathcal{P}}_s^{tm}(0)> \end{array}\r) \, \label{eq:corr_matrix}
\end{equation}
is build from operators $\tilde{\mathcal{S}}_l^{3,tm}$, $\tilde{\mathcal{P}}_s^{tm}$ that are
renormalized up to $Z_P$ and apart from that correspond to the ones in
Eqs.~(\ref{eq:light_S_eta_operator},\ref{eq:heavy_P_tm_eta_operator_renormalized}).

When building this matrix, we subtract the excited states in the connected contributions and apply the
required factors of $Z_S/Z_P$, as given in Tab.~\ref{tab:tabNf211}. We ignore the global factor $Z_P^2$
as it will cancel analytically in all observables. Then, we proceed by solving the generalized eigenvalue
problem \cite{Michael:1982gb,Luscher:1990ck,Blossier:2009kd}:
\begin{equation}
 \tilde{\mathcal{C}}\l(t\r) v^{\l(n\r)}\l(t,t_0\r) = \lambda^{\l(n\r)}\l(t,t_0\r)
\tilde{\mathcal{C}}\l(t_0\r) v^{\l(n\r)}\l(t,t_0\r) \,, \qquad t_0<t \,, \label{eq:GEVP}
\end{equation}
which gives access to masses through
\begin{equation}
 \frac{\lambda^{(n)}(t,t_0)}{\lambda^{(n)}(t+1,t_0)} = \frac{\exp\l(-m^{\l(n\r)}t\r)+
\exp\l(-m^{\l(n\r)}\l(T-t\r)\r)}{\exp\l(-m^{\l(n\r)}\l(t+1\r)\r) +
\exp\l(-m^{\l(n\r)}\l(T-\l(t+1\r)\r)\r)} \, \label{eq:masses}
\end{equation}
and amplitudes
\begin{equation}
 A^{\l(n\r)}_i = \frac{\sum\limits_{j=1}^N\tilde{\mathcal{C}}_{ij}\l(t\r)
v^{\l(n\r)}_j\l(t,t_0\r)}{\sqrt{\l(v^{\l(n\r)}\l(t,t_0\r), \mathcal{C}\l(t\r) v^{\l(n\r)}\l(t,t_0\r)\r)
\l(\exp\l(-m^{\l(n\r)} t\r) \pm \exp\l(m^{\l(n\r)} \l(t-T\r)\r)\r)}} \,, \label{eq:amplitudes}
\end{equation}
respectively. Since masses are renormalization group invariants, they are not affected by the actual
choice of $Z_S/Z_P$, which is only relevant for the computation of amplitudes leading to different
lattice artifacts for the methods M1 and M2. Note that the amplitudes computed in this way are only
renormalized up to a factor of $Z_P$ as well.

In Tab.~\ref{tab:setup_nf211}, we give the number of configurations $N$ used on each gauge ensemble to
extract observables in the flavor singlet sector. Errors are computed using bootstrapping with 1000
samples and we have used blocking to deal with autocorrelations. The numbers of configurations per block
$N_b$ are given in Tab.~\ref{tab:setup_nf211} and have been chosen to correspond to a block length of at
least 20 HMC trajectories. In addition, we give the number of stochastic samples that has been employed
in the computation of quark disconnected diagrams. It has been chosen such that the resulting statistical
errors are dominated by gauge noise.

The results for all ensembles are given in Tab.~\ref{tab:masses}. Clearly, the statistical error for a
test of the Witten-Veneziano formula will be dominated by flavor singlet related quantities as can be
inferred from the relative errors on the masses. This is caused by large contributions from quark
disconnected diagrams in the flavor singlet sector which are intrinsically more noisy compared to
connected contributions and also introduce sizable autocorrelations in case of the $\eta'$ mass.

For the computation of the kaon masses, we refer to Ref.~\cite{Baron:2010bv,Ottnad:2012fv}. The values
are also listed in Tab.~\ref{tab:masses}, together with the corresponding charged pion masses $M_{\rm
PS}$. Note that some of these values were calculated at smaller statistics than the flavor singlet sector
(c.f. Tab.~\ref{tab:setup_nf211}). Nevertheless, their relative statistical error is still at least one
order of magnitude smaller than for quantities in the flavor singlet sector.

\begin{table}[t]
 \centering
 \begin{tabular*}{1.0\textwidth}{@{\extracolsep{\fill}}lcccc}
  \hline\hline
  ensemble & $aM_{\mathrm{PS}}$ & $aM_\mathrm{K}$ & $aM_\eta$ & $a M_{\eta'}$ \\
  \hline\hline
  $A30.32$   & 0.12358(30) & 0.25150(29) & 0.2800(55) & 0.480(17) \\
  $A40.24$   & 0.14484(44) & 0.25884(43) & 0.2834(35) & 0.427(14) \\
  $A40.32$   & 0.14140(30) & 0.25666(23) & 0.2809(28) & 0.458(22) \\
  $A60.24$   & 0.17277(48) & 0.26695(52) & 0.2908(46) & 0.471(16) \\
  $A80.24$   & 0.19870(35) & 0.27706(61) & 0.3004(17) & 0.479(23) \\
  $A100.24$  & 0.22127(32) & 0.28807(34) & 0.3063(24) & 0.454(15) \\
  \hline
  $A80.24s$  & 0.19822(33) & 0.25503(33) & 0.2686(30) & 0.463(14) \\
  $A100.24s$ & 0.22118(33) & 0.26490(74) & 0.2763(13) & 0.518(30) \\
  \hline
  $B25.32$   & 0.10685(43) & 0.21240(50) & 0.2348(41) & 0.414(19) \\ 
  $B35.32$   & 0.12496(45) & 0.21840(28) & 0.2363(23) & 0.435(24) \\ 
  $B55.32$   & 0.15396(31) & 0.22799(34) & 0.2469(28) & 0.480(26) \\ 
  $B75.32$   & 0.18036(39) & 0.23753(32) & 0.2537(23) & 0.430(23) \\ 
  $B85.24$   & 0.19373(64) & 0.24392(59) & 0.2640(28) & 0.453(28) \\
  \hline  
  $D15.48$   & 0.06954(26) & 0.16897(85) & 0.1891(64) & 0.295(21) \\
  $D30.48$   & 0.09801(25) & 0.17760(23) & 0.1969(70) & 0.288(21) \\
  $D45.32sc$ & 0.11991(37) & 0.17570(84) & 0.1898(22) & 0.276(15) \\
  \hline\hline
 \end{tabular*}
 \caption{Results for meson masses on all dynamical ensembles
\cite{Baron:2010bv,Ottnad:2012fv,Michael:2013gka}. In addition to the relevant masses for the
Witten-Veneziano formula in Eq.~(\ref{eq:W-Vformula}), we also give the corresponding values of the
charged pion mass $M_\mathrm{PS}$.}
 \label{tab:masses}
\end{table}

\subsection{Treatment of \texorpdfstring{$f_0$}{f0}}
Consider the general definition for the decay constant $f_P$, \footnote{The normalization of decay
constants in this work has been chosen consistently s.t. $f_{\rm PS} \approx 130 \,\mathrm{MeV}$ holds
for the decay constant of the charged pion.} of any pseudoscalar meson $P$:
\begin{equation}
 \l< 0 \r| A_\mu^a\l(0\r) \l| P\l(p\r) \r> = i f^a_P p_\mu \,,
 \label{eq:matrix_element_A}
\end{equation}
where $A_\mu^a\l(0\r)$ denotes the axial vector current with flavor structure denoted by the index $a$.

In the charged sector for mass-degenerate, light quarks, the axial vector current in the physical basis
transforms into the vector current in the twisted basis at maximal twist. This feature can be exploited
together with the PCVC relation to derive an expression for the charged pion decay constant
$f_\mathrm{PS}$ \cite{Frezzotti:2001du,DellaMorte:2001tu,Jansen:2003ir}:
\begin{equation}
f_\mathrm{PS} = 2 \mu_l \frac{\bra{0} P^a_l \ket{\pi^\pm}}{M_\mathrm{PS}^2} \,,
\label{eq:f_PS}
\end{equation}
where $a=1,2$. Since this formula depends on a pseudoscalar matrix element, it can be calculated with
very high statistical accuracy. Moreover, this relation allows one to compute $f_{\rm PS}$ in the twisted
mass formulation without the need for any renormalization factor at all, because a factor $Z_P^{-1}$ for
the bare quark mass $\mu_l$ cancels the factor for the pseudoscalar matrix element in the twisted basis.

Similar considerations apply for the kaon sector, although a complication arises from the fact that one
has to employ interpolating operators made of light and heavy quarks. The relation for the kaon decay
constant $f_K$ in the twisted mass formulation is given by:
\begin{equation}
 f_K = \l(\mu_l + \mu_s \r)\frac{\bra{0} \tilde{\mathcal{P}}^{+,tm}_{neutral} \ket{K}}{M_K^2} \,,
 \label{eq:f_K}
\end{equation}
where $\tilde{\mathcal{P}}^{+,tm}_{neutral} = \frac{1}{2}( \frac{Z_S}{Z_P}(-\bar{\chi}_d\chi_c + \bar{\chi}_d\chi_s) +
\bar{\chi}_d i\g{5} \chi_c + \bar{\chi}_d i\g{5} \chi_s) $. In this case the factor
$Z_S/Z_P$ is required again, because it enters $\mu_s$ (c.f. Eq.~(\ref{eq:heavy_quark_masses})) as well
as relative factors due to mixing between scalar and pseudoscalar currents.

Assuming exact isospin symmetry and neglecting possible contributions from the charm quark and mixing
with further states such as glueballs, the most general parametrization for decay constants of the
$\eta$, $\eta'$ system reads
\begin{equation}
 \l( \begin{array}{ll}
   f_\eta^8 & f_\eta^0 \\
   f_{\eta'}^8 & f_{\eta'}^0 
  \end{array}\r) = \l( \begin{array}{rr}
   f_8 \cos \phi_8  & -f_0 \sin \phi_0 \\
   f_8 \sin \phi_8 & f_0 \cos \phi_0
  \end{array} \r) \equiv \Xi\l(\phi_8, \phi_0\r) \diag \l(f_8,\, f_0\r) \,.
  \label{eq:singlet_octet_basis_parametrization}
\end{equation}
The choice of $A^0_\mu$, $A^8_\mu$ together with $\mathrm{P}=\eta,\eta'$ in
Eq.~(\ref{eq:matrix_element_A}) defines the so-called \emph{singlet-octet basis}.

Moreover, employing $\chi$PT, it is possible to relate the decay constant parameters in the
$\eta$--$\eta'$ system to the remaining octet decay constants $f_\pi$, $f_K$ and a low energy constant
$\Lambda_1=\mathcal{O}(1/N_c)$ occurring at next-to-leading order in the chiral
expansion~\cite{Kaiser:1998ds,Feldmann:1998vh,Feldmann:1999uf} by:
\begin{alignat}{2}
 f_0^2   &= \bigl(f_\eta^0\bigr)^2 + \bigl(f_{\eta'}^0\bigr)^2 &&= \frac{1}{3} \l(2 f_K^2 + f_\pi^2 \r) + \Lambda_1 f_\pi^2  \,, \label{eq:f_0_f_8_f_pi_f_K_relations_1} \\
 f_8^2   &= \bigl(f_\eta^8\bigr)^2 + \bigl(f_{\eta'}^8\bigr)^2 &&= \frac{1}{3} \l(4 f_K^2 - f_\pi^2 \r) \,, \label{eq:f_0_f_8_f_pi_f_K_relations_2} \\
 f_0 f_8 \sin\bigl(\phi_8 - \phi_0\bigr) &= f_\eta^0 f_\eta^8 + f_{\eta'}^0 f_{\eta'}^8 &&= - \frac{2\sqrt{2}}{3} \l(f_K^2 - f_\pi^2 \r) \,. \label{eq:f_0_f_8_f_pi_f_K_relations_3}
\end{alignat}
In the chosen basis and to the given chiral order, the additional, OZI-violating corrections are specific
to the singlet sector, i.e. the term $\sim \Lambda_1 = \order{N_c^{-1}}$ affects neither $f_8$, nor any
of the angles $\phi_0$, $\phi_8$, but only the parameter $f_0$. From the last relation, it can be
inferred that in the octet singlet basis, the difference between the two angles $\phi_0$, $\phi$ is given
by $\mathrm{SU}(3)_F$--violating effects, leading to the expectation
\begin{equation}
 \l| \frac{\phi_0 - \phi_8}{\phi_0 + \phi_8}\r| \slashed{\ll} 1 \,.
\end{equation}

The fact that there are different types of contributions to the mixing, i.e. $\SU{3}_F$-breaking and
OZI-violating effects $\sim \Lambda_1$, can be exploited in order to choose a basis in which the two
resulting mixing angles do not exhibit a sizable splitting. To this end, one introduces the \emph{quark
flavor basis} with the axial vector currents $A_\mu^0$ and $A_\mu^8$ replaced by the combinations
\begin{alignat}{2}
 A^l_\mu =& \frac{2}{\sqrt{3}} A^0_\mu + \sqrt{\frac{2}{3}} A^8_\mu =&& \frac{1}{\sqrt{2}} \l(\bar{u} \g{\mu} \g{5} u + \bar{d} \g{\mu} \g{5} d\r) \,, \label{eq:quark_flavor_basis_1} \\ 
 A^s_\mu =& \sqrt{\frac{2}{3}} A^0_\mu - \frac{2}{\sqrt{3}} A^8_\mu =&& \bar{s} \g{\mu} \g{5} s \,, \label{eq:quark_flavor_basis_2}
\end{alignat}
in which the light quarks and the strange quark contributions are disentangled. This is the reason why
this basis is more convenient in lattice simulations, as this is also the preferred flavor structure for
interpolating operators on the lattice, allowing one to directly access the corresponding matrix
elements. In exact analogy to the singlet-octet basis, this basis again allows for a parametrization in
terms of two decay constants and two mixing angles
\begin{equation}
 \l( \begin{array}{ll}
 f_\eta^l & f_\eta^s \\
 f_{\eta'}^l & f_{\eta'}^s 
 \end{array}\r) = \Xi\l(\phi_l,\phi_s\r)  \diag\l(f_l,\, f_s\r) \,,
 \label{eq:quark_flavor_basis_parametrization}
\end{equation}
where the mixing matrix $\Xi$ has the same form as the one defined in
Eq.~(\ref{eq:singlet_octet_basis_parametrization}). In this basis, the relations between mixing
parameters in the $\eta$--$\eta'$ system and $f_\pi$, $f_K$, $\Lambda_1$ read:
\begin{alignat}{2}
 f_l^2   &= \bigl(f_\eta^l\bigr)^2 + \bigl(f_{\eta'}^l\bigr)^2 &&= f_\pi^2 + \frac{2}{3} \Lambda_1 f_\pi^2  \,, \label{eq:f_l_f_s_f_pi_f_K_relations_1} \\
 f_s^2   &= \bigl(f_\eta^s\bigr)^2 + \bigl(f_{\eta'}^s\bigr)^2 &&= 2 f_K^2 - f_\pi^2 + \frac{1}{3} \Lambda_1 f_\pi^2 \,, \label{eq:f_l_f_s_f_pi_f_K_relations_2} \\
 f_l f_s \sin\bigl(\phi_l - \phi_s\bigr) &= f_\eta^l f_\eta^s + f_{\eta'}^l f_{\eta'}^s &&= \frac{\sqrt{2}}{3} \Lambda_1 f_\pi^2 \,. \label{eq:f_l_f_s_f_pi_f_K_relations_3}
\end{alignat}
The most important feature of the quark flavor basis becomes manifest in the last expression, which is
now entirely given by an OZI-violating contribution $\sim\Lambda_1=\order{\mathcal{\delta}}$, amounting to
additional suppression for the difference $\l|\phi_l-\phi_s\r|$ compared to $\l|\phi_0-\phi_8\r|$, which
is given by $\mathrm{SU}(3)_F$--breaking effects. Besides, in the $\SU{3}_F$ symmetric case, the angles
$\phi_l \approx \phi_s$ take the value $\phi_{\SU{3}_F} = \arctan\sqrt{2}$, and, hence, their numerical
value is not expected to be small. Therefore, one expects:
\begin{equation}
 \l| \frac{\phi_l - \phi_s}{\phi_l + \phi_s}\r| \ll 1 \,
\end{equation}
in the quark flavor basis, which has been numerically confirmed in a previous lattice study
\cite{Michael:2013gka}. This feature allows one to consider a simplified mixing scheme in the quark
flavor basis with only one angle $\phi$
\begin{equation}
 \l( \begin{array}{ll}
 f_\eta^l & f_\eta^s \\
 f_{\eta'}^l & f_{\eta'}^s 
 \end{array}\r) = \Xi\l(\phi\r) \diag\l(f_l,\, f_s\r) + \order{\Lambda_1} \,,
 \label{eq:simplified_mixing_scheme}
\end{equation}
where $\Xi\l(\phi\r) \equiv \Xi\l(\phi, \phi\r)$. 

As mentioned at the beginning of this section, we are restricted in our simulations to pseudoscalar
operators for practical purposes, because the signal-to-noise ratio for axial vector operators turns out
too small for a direct computation of the relevant observables. However, one may instead consider
pseudoscalar matrix elements in order to retrieve information on the mixing parameters. This is possible
due to the relation between axial vector and pseudoscalar matrix elements, which is given
non-perturbatively by 
\begin{equation}
 \partial^\mu A_\mu^a = \bar{\psi}\l(x\r)  2 M T^a i\g{5} \psi\l(x\r) + \delta^{0a} \sqrt{2 N_f} \, \omega\l(x\r) \,,
\end{equation}
where for $a=0$ we have $T^0 = \sqrt{1/(2 N_f)}\,\mathds{1}_{N_f \times N_f}$ and $\omega\l(x\r)$ denotes the winding
number density.

Nonetheless, the anomaly equation of QCD itself is not sufficient for any practical purposes here,
mainly because it requires knowledge of an additional matrix element involving the topological charge
density. Therefore, one needs to gain further insight on how the pseudoscalar matrix elements are linked
to the mixing parameters. This is again achieved by the use of $\chi$PT. Consider pseudoscalar currents
in the quark flavor basis in analogy to
Eqs.~(\ref{eq:quark_flavor_basis_1}),(\ref{eq:quark_flavor_basis_2}):
\begin{align}
 P^l &= \frac{1}{\sqrt{2}} \l( \bar{u} i\g{5} u + \bar{d} i\g{5} d \r) \,, \label{eq:pseudoscalar_quark_flavor_basis_1} \\
 P^s &= \bar{s} i\g{5} s \,, \label{eq:pseudoscalar_quark_flavor_basis_2}
\end{align}
and the corresponding matrix elements for pseudoscalar mesons $\mathrm{P}$ that are given by
\begin{equation}
 h_\mathrm{P}^i = 2 m_i \l<0\r| P^i \l|\mathrm{P}\r> \,.
 \label{eq:pseudoscalar_matrix_elements}
\end{equation}
To leading order, one can make contact with the quark flavor basis parametrization for axial vector
matrix elements in Eq.~(\ref{eq:quark_flavor_basis_parametrization}), i.e. obtain an expression for
$h_\mathrm{P}^i$ in terms of decay constants $f_l$, $f_s$, the mixing angle $\phi$ and octet meson masses
\cite{Feldmann:1999uf}:
\begin{equation}
 \l( \begin{array}{ll}
 h_\eta^l & h_\eta^s \\
 h_{\eta'}^l & h_{\eta'}^s 
 \end{array}\r) = \Xi\l(\phi\r) \diag\l(M_\pi^2 f_l,\, \l(2 M_\mathrm{K}^2 - M_\pi^2\r) f_s\r) \,.
 \label{eq:pseudoscalar_quark_flavor_basis_parametrization}
\end{equation}
Again, formally higher order, OZI-violating contributions are neglected in this expression, as demanded
by the so-called FKS-scheme, which allows for the determination of process-independent mixing parameters
\cite{Feldmann:1998vh,Feldmann:1998sh,Feldmann:1999uf}.

Finally, we can combine
Eqs.~(\ref{eq:f_0_f_8_f_pi_f_K_relations_1},\ref{eq:f_0_f_8_f_pi_f_K_relations_2}) and
Eqs.~(\ref{eq:f_l_f_s_f_pi_f_K_relations_1},\ref{eq:f_l_f_s_f_pi_f_K_relations_2}) to write down leading
order relations for the desired singlet decay constant $f_0$ in terms of the parameters $f_\mathrm{PS}$,
$f_K$, $f_l$ and $f_s$, which we compute on the lattice:
\begin{align}
 f_0^2 &= - 7/6 f_{\rm PS}^2 + 2/3 f_K^2 + 3/2 f_l^2 \,,     \label{eq:f_0_def1} \\
 f_0^2 &= + 1/3 f_{\rm PS}^2 - 4/3 f_K^2 + f_l^2 + f_s^2 \,, \label{eq:f_0_def2} \\
 f_0^2 &= + 8/3 f_{\rm PS}^2 -16/3 f_K^2 + 3 f_s^2 \,.       \label{eq:f_0_def3}
\end{align}
 
In general, these relations receive corrections of $\mathcal{O}(\delta^2)$. However, since they were
derived from continuum $\chi$PT, they also differ by lattice artifacts of $\mathcal{O}(a^2)$, if applied
to our lattice data. We will in the following exploit this ambiguity to choose a definition for $f_0$
that exhibits particularly small systematic effects. Although the use of pseudoscalar matrix elements and
changing the flavor basis requires to resort to $\chi$PT, we point out that this is not a serious
drawback, as the above expressions are in general of the same order in the chiral expansion as the
Witten-Veneziano formula in Eq.~(\ref{eq:W-Vformula}). In the following, we will refer to the three
definitions Eqs.~(\ref{eq:f_0_def1}--\ref{eq:f_0_def3}) of $f_0$ as D1, D2 and D3, respectively.

Regarding the computation of decay constants in the charged meson sector, we point out that while the
values of $f_\mathrm{PS}$ have been recalculated with the current statistics (they were first published
in \cite{Baron:2010bv}), we used less configurations for $f_K$. However, by far the largest contribution
to the overall statistical error in our analysis stems again from the flavor singlet sector. Finally, we
remark that a dedicated study of $\eta$,$\eta'$--related decay constant parameters is currently in
preparation \cite{decayconst}.


\subsection{Chiral extrapolations}
Since our lattice simulations employ unphysical quark masses, we need to perform chiral extrapolations of
our lattice data when computing the l.h.s. of Eq.~(\ref{eq:W-Vformula}). In principle, to the given order
in $\chi$PT, this simply corresponds to a constant fit in $(r_0M_\mathrm{PS})^2$. However, we have to
take into account lattice artifacts which might be different for the three definitions
Eqs.~(\ref{eq:f_0_def1}--\ref{eq:f_0_def3}). 

Moreover, the dependence on the choice of $Z_P/Z_S$ is non-trivial, as it affects the relevant decay
constants (besides $f_\mathrm{PS}$) directly through the renormalization of the corresponding matrix
elements, as well as through the relative renormalization factor that enters the quark mass $\mu_s$ in
Eq.~(\ref{eq:heavy_quark_masses}), which appears in the definition of $f_K$ in Eq.~(\ref{eq:f_K}).

\begin{figure}[!ht]
 \begin{center}
  \includegraphics[width=0.49\textwidth]{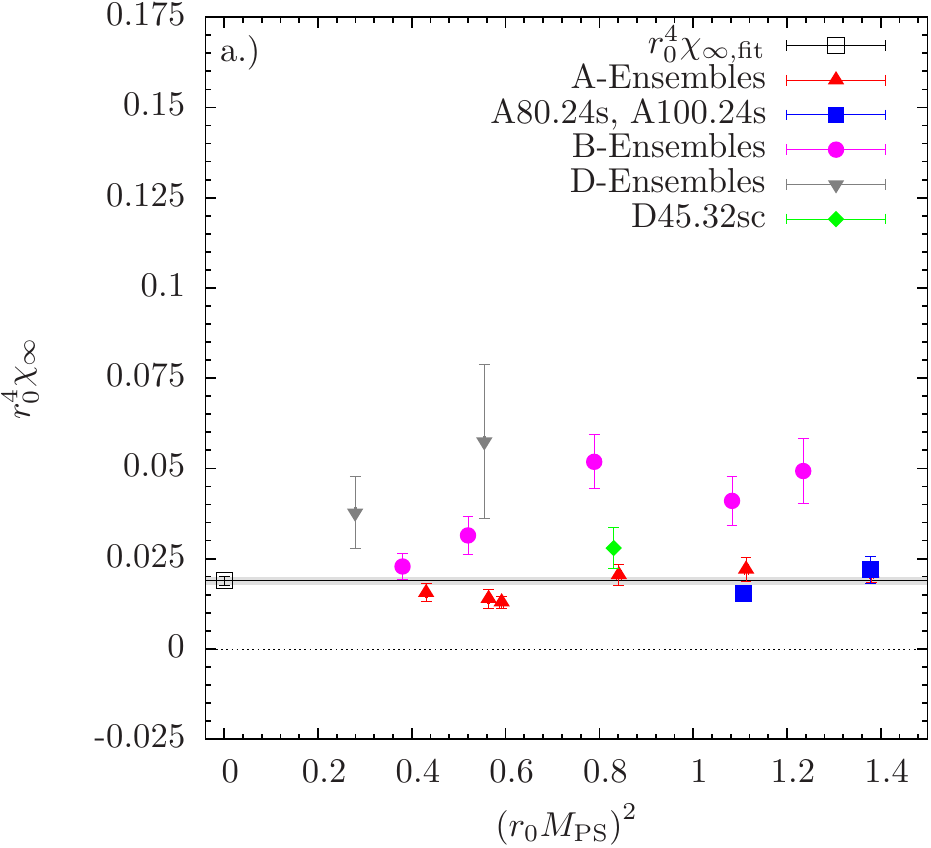}
  \includegraphics[width=0.49\textwidth]{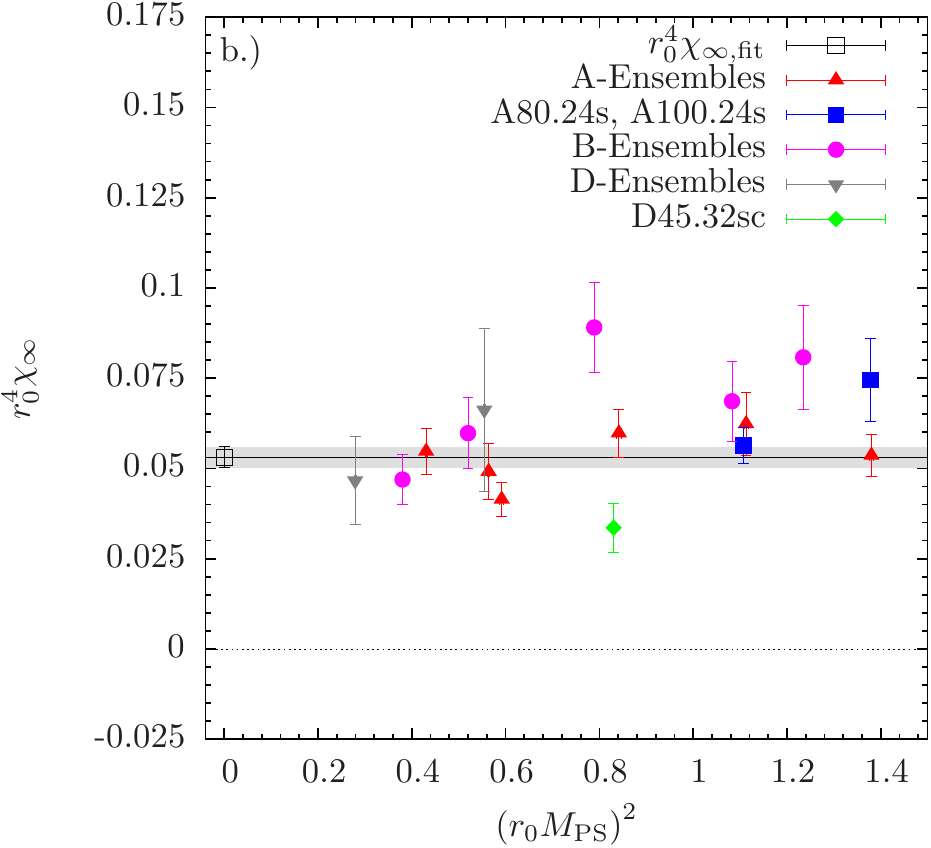}\\
  \includegraphics[width=0.49\textwidth]{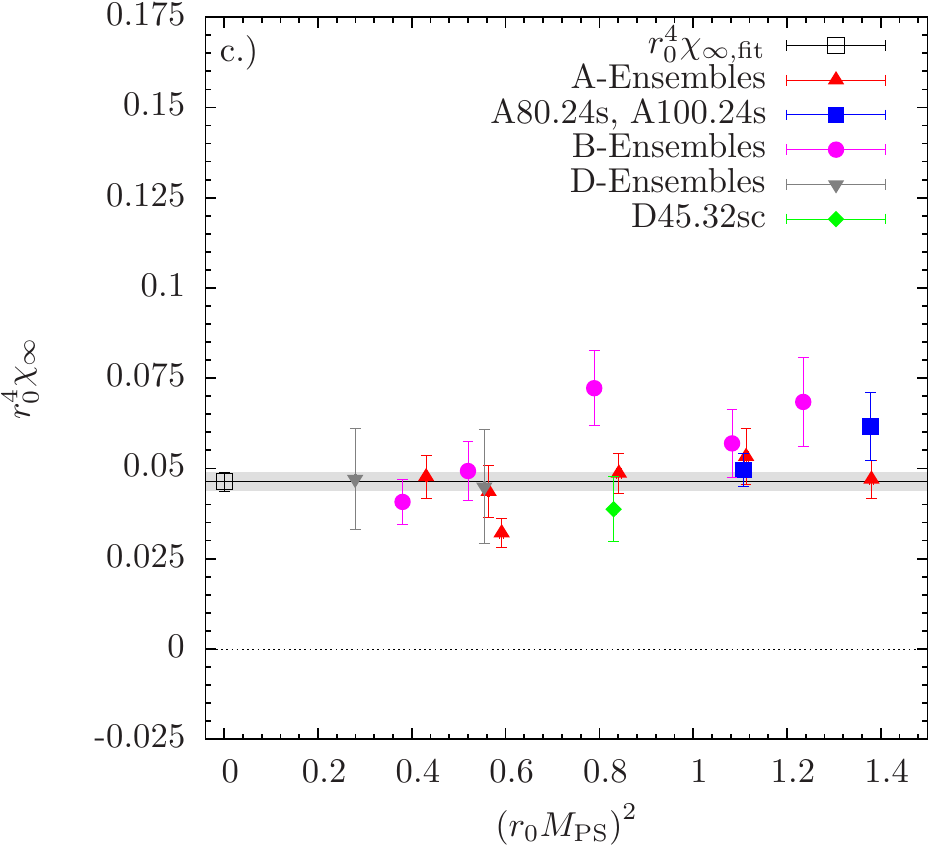}
  \includegraphics[width=0.49\textwidth]{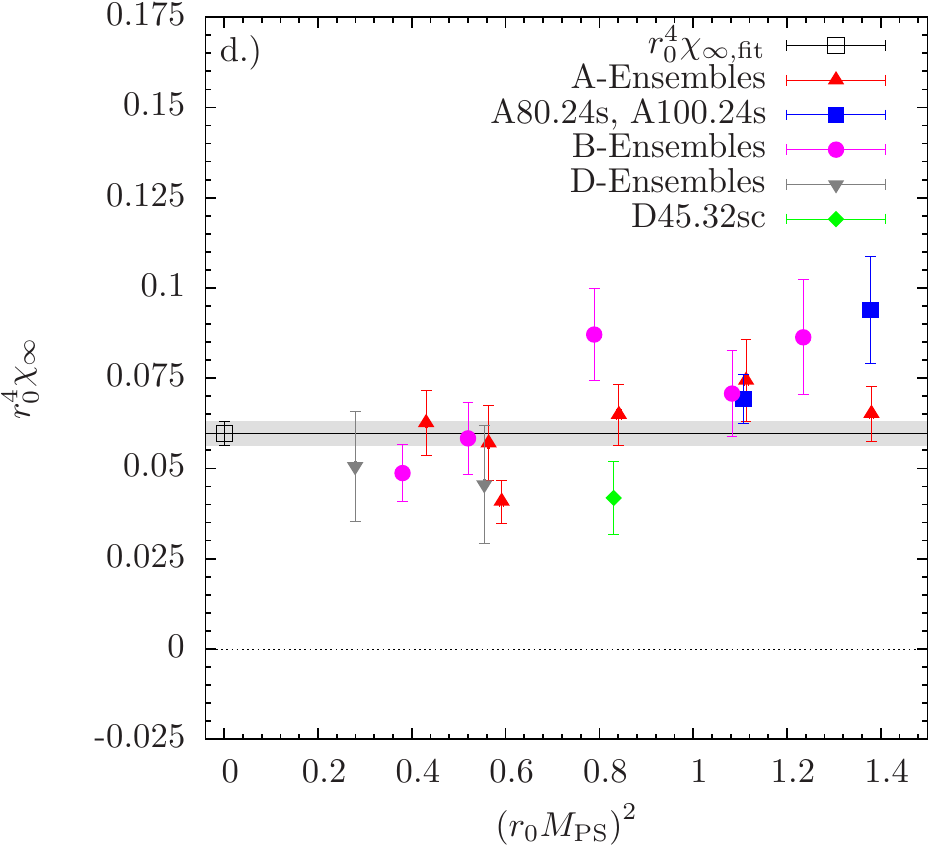}\\
  \includegraphics[width=0.49\textwidth]{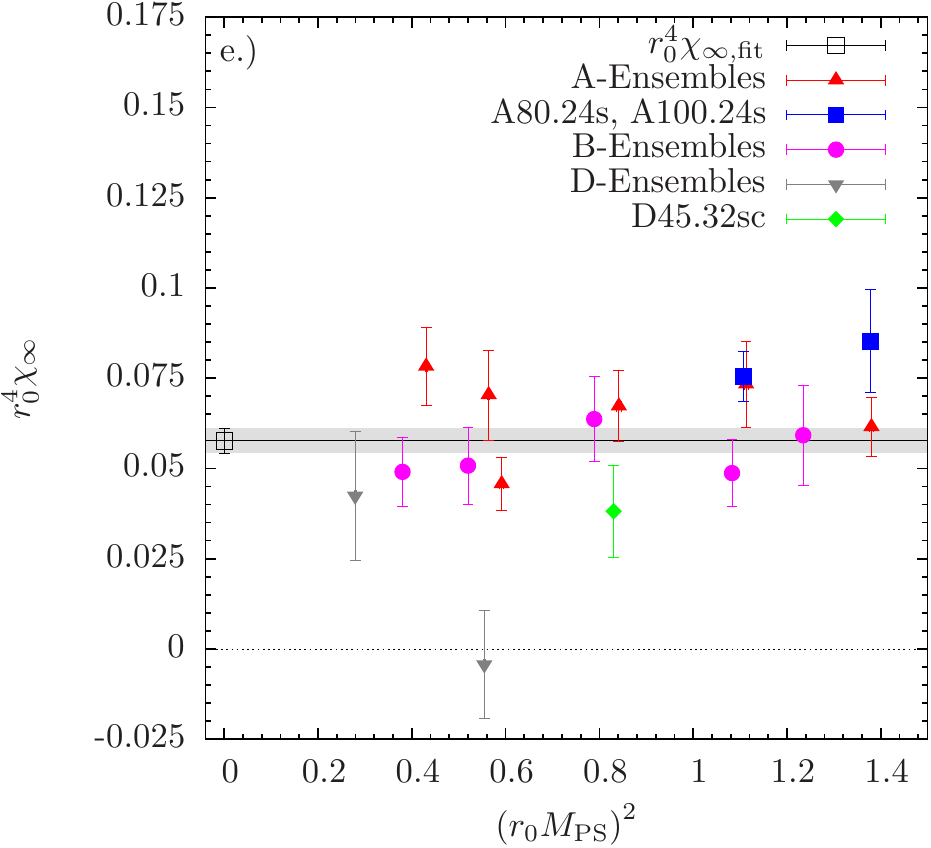}
  \includegraphics[width=0.49\textwidth]{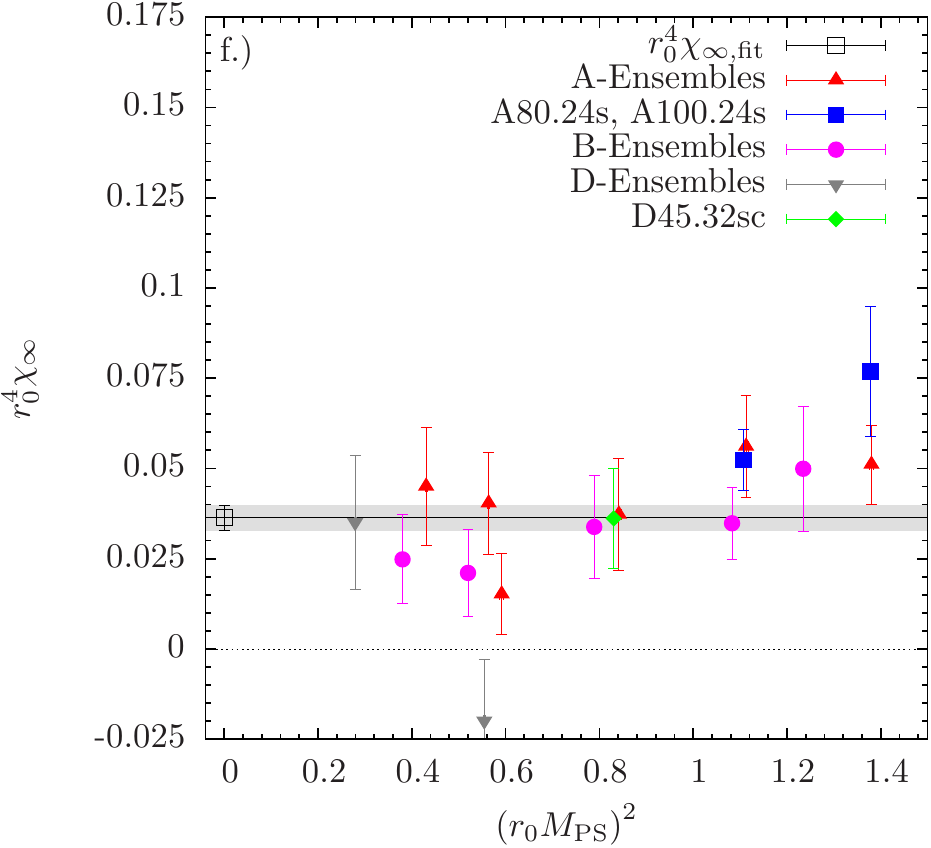}
 \end{center}
 \vspace{-0.75cm}
 \caption{\label{fig:chi_fullQCD} Results for $r_0^4\chi_\infty$ calculated from meson masses and $f_0$
as a function of $(r_0 M_{\rm PS})^2$. The values of $f_0$ employed in panels a.), b.) are obtained from
Eq.~(\ref{eq:f_0_def1}), for c.), d.) from Eq.~(\ref{eq:f_0_def2} and for e.), f.) from
Eq.~(\ref{eq:f_0_def3}). The plots in the left and right column correspond to different $Z_P/Z_S$ values
from M1 and M2, respectively, as listed in Tab.~\ref{tab:tabNf211}. The chirally extrapolated value is
obtained from a constant fit in $(r_0 M_{\rm PS})^2$ and has to be compared with the value computed in
the pure Yang-Mills theory.}
\end{figure}

In Fig.~\ref{fig:chi_fullQCD} we show results for $r_0^4\chi_\infty$ calculated from using our results
for the relevant meson masses as given in Tab.~\ref{tab:masses}. For the plots a.), c.) and e.) in the
left column we used $Z_P/Z_S$ values from method M1 and for the right ones b.), d.) and f.) those from
M2; c.f. Tab.~\ref{tab:tabNf211}.
The three rows in Fig.~\ref{fig:chi_fullQCD} correspond to the three definitions of $f_0$ D1, D2 and D3,
respectively.

Clearly, the three definitions of $f_0$ show different lattice artifacts and systematic effects regarding
their $Z_P/Z_S$ (and hence $m_s$) dependence. Besides these systematic effects, their relative
statistical errors differ as well. Within errors, the first definition D1 shows the largest lattice
artifacts, as well as the most significant dependence on the choice of $Z_P/Z_S$. Applying a constant
fit in $(r_0 M_\mathrm{PS})^2$ does not provide a good description of the data in this case, as can be
seen from panel a.). The fitted value is much lower than for any other definition and most points are
incompatible with the fitted line. However, the data points show a trend towards larger values at smaller
lattice spacings, i.e. by fitting only to the data at the finest lattice spacing value we obtain
$r_0^4\chi_\infty^D=0.043(12)_\stat$, significantly higher than the result of the fit to all data points.

The data points extracted from definitions D2 and D3 lead to a reasonable agreement with a constant
extrapolation in $(r_0M_\mathrm{PS})^2$, with clearly the best fits stemming from definition D3.
In particular, with definition D3, the extrapolated value is merely independent on the choice of $Z$.

We have also tried to add a term of $\mathcal{O}(a^2)$ to our fit. It turns out that apart from the most
extreme case shown in panel a.) of Fig.~\ref{fig:chi_fullQCD}, the data does not allow to resolve such a
dependence within errors, i.e. the results for the corresponding coefficient are compatible with zero.
Regarding the dependence on the strange quark mass, we point out that to the given chiral order, the
computed $\chi_\infty$ is expected to be a constant function of $(r_0 M_K)^2$ as well. Although the data
shows a residual dependence on the strange quark mass, no clear picture arises with respect to its
functional form.

We have collected numerical results from our constant fits for $\chi_\infty$ in Tab.~\ref{tab:fits} for
the three definitions of $f_0$ and the two sets of values for $Z_P/Z_S$. In addition, we give the
respective $\chi^2/\mathrm{dof}$ values. Clearly, the fit M1D1 is by far worst, which is expected due to
the large cutoff effects in this case.

\begin{table}[t]
 \centering
 \begin{tabular*}{1.0\textwidth}{@{\extracolsep{\fill}}lcccccc}
  \hline\hline
  method & M1D1 & M1D2 & M1D3 & M2D1 & M2D2 & M2D3 \\
  \hline\hline
  $r_0^4\chi_\infty$   & 0.019(1) & 0.046(3) & 0.058(3) & 0.053(3) & 0.060(3) & 0.036(4) \\
  $\chi^2/\mathrm{dof}$ & 4.90     & 1.71     & 2.60     & 2.19     & 2.04     & 1.95 \\
  \hline\hline
 \end{tabular*}
 \caption{Results from a constant fit for $r_0^4\chi_\infty$ and
   corresponding $\chi^2/\mathrm{dof}$ values ($\mathrm{dof}=15$) for two different sets of values of
$Z_P/Z_S$ (M1,M2) and three definitions for $f_0$ (D1,D2,D3); c.f.
Eqs.~(\ref{eq:f_0_def1}-\ref{eq:f_0_def3}). Errors are statistical only.}
 \label{tab:fits}
\end{table}

In order to obtain our final result for $\chi_\infty$ from the dynamic simulations, we apply a weight to
each fit:
\begin{equation}
w = 1 - 2|p - 0.5| \,,
\end{equation}
where $p$ denotes the $p$--values corresponding to the values of $\chi^2/\mathrm{dof}$ given in
Tab.~\ref{tab:fits} and take the average over all six fits, leading to:
\begin{equation}
  r_0^4\chi_\infty=0.047(3)_\stat(11)_\sys\,. \nonumber
\end{equation}
The systematic error has been chosen as the mean absolute deviation from the central value and should
reflect the uncertainties from residual cutoff and strange quark mass effects. Since we included the fit
M1D1, which suffers from particularly large lattice artifacts, the value of the systematic error should
be considered a conservative estimate.

Another possibility to deal with the residual effects of quark mass dependence and the lattice spacing is to include additional, higher order terms in the fit function
\begin{equation}
 f\l(r_0^4 \chi_\infty, (r_0 M_\pi)^2, (r_0 M_K)^2, (a/r_0)^2\r) = r_0^4\chi_\infty + c_1 (r_0 M_\pi)^2 + c_2 (r_0 M_K)^2+ c_3 (a/r_0)^2\,,
 \label{eq:lin_fit}
\end{equation}
where $c_i$ with $i=1,2,3$ denotes the free fit parameters. Note that from the point of view of chiral perturbation theory including only a subset of the linear terms would be inconsistent with power counting. As can be inferred from table~\ref{tab:linear_fits} this fit model leads to improved $\chi^2/\mathrm{dof}$ values. However, most of the additional terms are poorly constrained by the data and are close or compatible with zero. In general, this leads to much larger statistical errors and all results for $r_0^4 \chi_\infty$ are compatible within errors. Taking the average of all six fits weighted by their respective p-value and statistical errors yields
\begin{equation}
 r_0^4\chi_\infty=0.051(24)_\stat \,, \nonumber
\end{equation}
which agrees with the previously computed result from constant fits. Since for the fit model in E.q.~(\ref{eq:lin_fit}) all fit results for $r_0^4 \chi_\infty$ are compatible with the averaged result within statistical errors, we refrain from quoting an additional systematic uncertainty, as it has been done for the constant fits.

We remark that the inclusion of even more complicated terms seems not feasible, as the linear terms in the above model are already rather poorly constrained by the data and close or even compatible with zero in many cases. Regarding the data point M1D1 it appears that the introduction of additional terms does still not lead to a good $\chi^2/\mathrm{dof}$ value for the corresponding fit, although the resulting value for $r_0^4 \chi_\infty$ agrees well with the other results. Therefore, we conclude that this data point is rather to be considered a statistical outlier than revealing any actual effect related to physics.

\begin{table}[t]
 \centering
 \begin{tabular*}{1.0\textwidth}{@{\extracolsep{\fill}}lcccccc}
  \hline\hline
  method & M1D1 & M1D2 & M1D3 & M2D1 & M2D2 & M2D3 \\
  \hline\hline
  $r_0^4\chi_\infty$ & 0.062(12) & 0.066(20) & 0.039(31) & 0.049(21) & 0.061(26) & 0.030(38) \\
  $\chi^2/\mathrm{dof}$ & 3.45 & 1.50 & 1.34 & 1.91 & 1.29 & 1.23 \\
  \hline\hline
 \end{tabular*}
 \caption{Results for $r_0^4\chi_\infty$ from the linear fit model in Eq.~(\ref{eq:lin_fit}) and the
 corresponding $\chi^2/\mathrm{dof}$ values ($\mathrm{dof}=12$) for two different sets of values of
 $Z_P/Z_S$ (M1,M2) and three definitions for $f_0$ (D1,D2,D3); c.f.
 Eqs.~(\ref{eq:f_0_def1}-\ref{eq:f_0_def3}). Errors are statistical only.}
 \label{tab:linear_fits}
\end{table}

\section{Discussion}

\begin{figure}[t]
  \centering
  \includegraphics{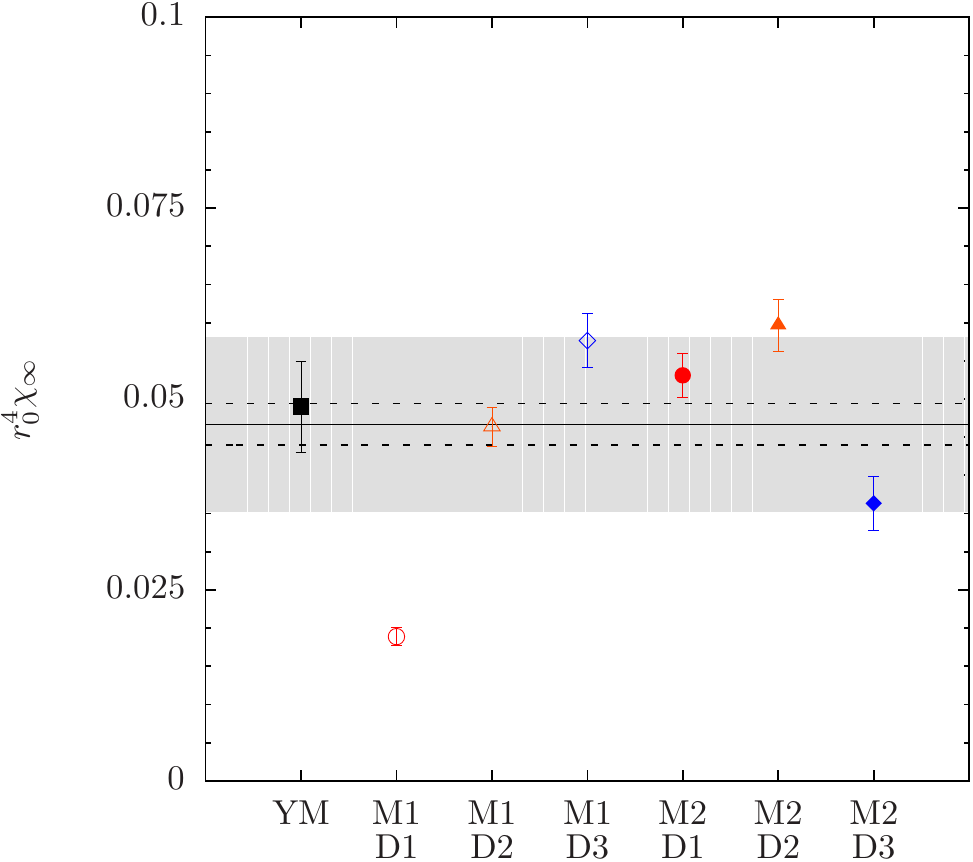}
  \caption{Results from the pure Yang-Mills theory (YM) and dynamical simulations. Open and closed
symbols correspond to the two sets of values for $Z_P/Z_S$ (M1, M2). All three definitions D1, D2 and D3
for $f_0$ are included and represented by circle, triangle and diamond symbols, respectively. The solid
black line represent the final, $p$--value weighted average from dynamical simulations and its
statistical error is indicated by the gray band. In addition, the dotted lines correspond to its
systematic error; see text.}
  \label{fig:comparison}
\end{figure}

In Fig.~\ref{fig:comparison}, we compare the pure YM topological susceptibility to the left hand
side of Eq.~(\ref{eq:W-Vformula}) computed in $N_f=2+1+1$ lattice QCD. All results are given in units of
the Sommer parameter $r_0$. 
The weighted mean of the dynamical results and its statistical error is indicated by the gray band.
The systematic uncertainty of the weighted average is shown by the dotted horizontal lines.
Apart from the outlier M1D1, which is affected by sizable cutoff effects, the results from the dynamical
simulations are very close to the one from the pure YM theory.
Also the agreement of the quenched and the averaged dynamical result appears to be good within
statistical and systematic uncertainties. 

A complication arises when one attempts to convert the results to physical units. 
It stems from the value of $r_0$ in the YM theory: while in $N_f=2+1+1$ QCD direct contact to physical
quantities is natural, in the YM theory such a relation is not obvious. 
In particular, the question arises whether or not the values of $r_0$ in the YM theory and in $N_f=2+1+1$
QCD are expected to be equal. From a general point of view, we do not see a reason for this to be true.

Therefore, one may use the standard value of $r_0=0.5\ \mathrm{fm}$ to convert the YM result to physical
units and obtain:
\begin{equation}
  \chi_\infty^\mathrm{YM}=(185.3(5.6)_{\rm stat+sys}\,{\rm MeV})^4\,. \nonumber
\end{equation}
For the dynamical simulations, on the other hand, we can use the value $r_0=0.474(14)\ \mathrm{fm}$ computed
in Ref.~\cite{Carrasco:2014cwa} to convert to physical units. 
Taking the aforementioned weighted average, we obtain:
\begin{equation}
  \chi_\infty^\mathrm{dyn}=(193.5(6.2)_\stat(13.4)_\sys\,{\rm MeV})^4\,, \nonumber
\end{equation}
where we have included the error on the physical value of $r_0$ in the statistical uncertainty. We find rather close agreement between quenched ($\chi_\infty^\mathrm{YM}$) and dynamical ($\chi_\infty^\mathrm{dyn}$) results with deviations of only up to
$\mathcal{O}(10\%)$.
This confirms the validity of the Witten-Veneziano formula to the
given order and for the assumed value of the quenched $r_0$ also in
physical units.

\section{Summary}

In this paper, we have presented a non-perturbative test of the famous Witten-Veneziano formula. 
This formula relates the large mass value obtained for the $\eta'$ meson to the anomalously broken axial
$U(1)$ symmetry in QCD.
It, therefore, provides important insights for our understanding of QCD and the generation of masses.

We have computed the topological susceptibility in the pure YM theory with dedicated quenched lattice
simulations using the so-called spectral projector method.
In particular, by using four values of the lattice spacing, we were able to perform a reliable continuum
extrapolation and, hence, control this major systematic uncertainty.

For the first time, we have computed the flavour singlet decay constant $f_0$ using $N_f=2+1+1$ lattice QCD.
Together with the $\eta$, $\eta'$ and kaon meson masses determined in Ref~\cite{Michael:2013gka}, this
allowed us to compute also the l.h.s. of the Witten-Veneziano formula.
Again, lattice artifacts are controlled by using three values of the lattice spacing. 
By using a wide range of light quark mass values, also the SU$(2)$ chiral extrapolation was performed in
a controlled way. 
The strange quark mass dependence was found to be not important for the Witten-Veneziano formula.

The comparison of the pure YM topological susceptibility and its
Witten-Veneziano counterpart from full QCD leads to agreement within errors.
This finding provides clear evidence for the hypothesis that the large mass of the $\eta'$ meson is due
the anomalously broken axial $U(1)$ symmetry in QCD.

\appendix
\section{Computation of \texorpdfstring{$\kappa_c$}{kappac}}
\label{sec:appendixA}

\begin{figure}[!ht]
 \begin{center}
 \includegraphics[width=0.9\textwidth]{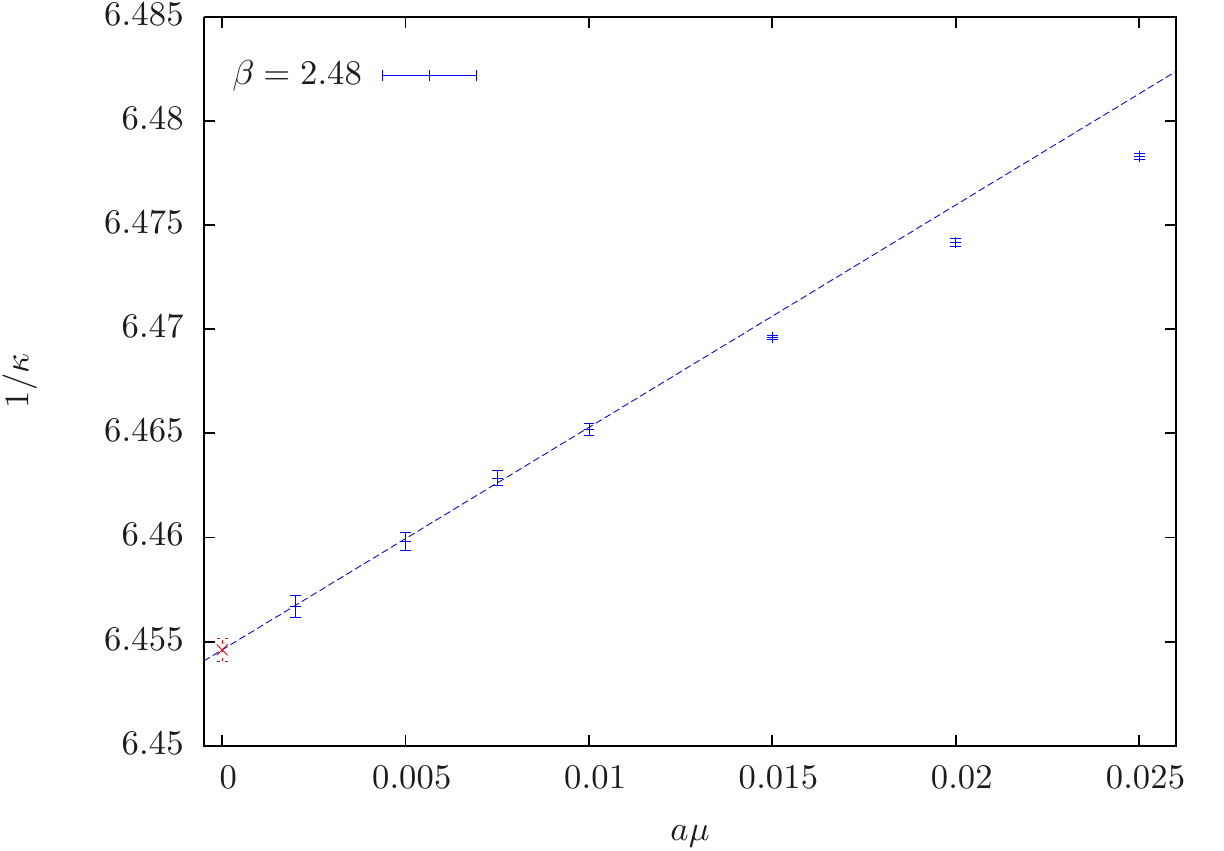}
 \caption[Chiral extrapolation of $\kappa_c$ for the quenched ensemble at
$\beta=2.48$]{\label{fig:chkappac}Chiral extrapolation of $\kappa_c$ for the quenched ensemble at
$\beta=2.48$. The red point corresponds to the linearly extrapolated value.}
 \end{center}
\end{figure}
     
In order to guarantee the $\obs(a)$ improvement, we need to tune $\kappa$ to its critical value for the
valence quarks. To do so, we followed the strategy introduced in \cite{Jansen:2005gf}. Thus, we computed
the of $\kappa$ through the evaluation of $m_{\rm pcac}$ for different values of the quark mass $a\mu$.
In particular, we imposed $m_{\rm pcac}<0.1a\mu$. In Fig.~\ref{fig:chkappac}, we show a particular
example of the chiral behavior of $\kappa_c$. In all cases, we perform a chiral fit considering  only the
lowest masses  $a\mu<0.01$ since the larger masses deviate  from the linear behavior. Notice that the
data is highly correlated and the fit needed to take this correlation into account.

\acknowledgments
We thank all members of ETMC for the most enjoyable collaboration. We are grateful 
to R. Frezzotti, U.-G.~Mei{\ss}ner and G.C. Rossi for helpful comments and discussion.
Furthermore, we would like to thank F. Zimmermann for contributions at an early stage of this project.
The computer time for this project was made available to us by the John von
Neumann-Institute for Computing (NIC) on the JUDGE and Jugene systems
in J{\"u}lich and the IDRIS (CNRS) computing center in Orsay. In particular 
we thank U.-G.~Mei{\ss}ner for granting us access on JUDGE. Further computational 
ressources were provided by SuperMUC at LRZ in Garching and the PC cluster in Zeuthen. 
This project was funded by the DFG as a project in the SFB/TR 16. K.C. has been 
supported in part by Foundation for Polish  Science fellowship ``Kolumb'' and by 
the Helmholtz International Center for FAIR within the framework of the LOEWE program 
launched by the State of Hessen. Two of the authors (K.O. and C.U.) were supported by 
the Bonn-Cologne Graduate  School (BCGS) of Physics and Astronomie. The open source software packages 
tmLQCD~\cite{Jansen:2009xp}, Lemon~\cite{Deuzeman:2011wz} and R~\cite{R:2005} have been used.

\bibliographystyle{JHEP}
\bibliography{refs}

\end{document}